\documentclass[printer]{aa}  
\usepackage{graphicx}
\usepackage{txfonts}
\usepackage{upgreek}
\usepackage{xcolor}
\usepackage[colorlinks=true, allcolors=blue]{hyperref}
\newcommand{\PRESTO}{\texttt{PRESTO}}
\newcommand{\TEMPO}{\texttt{TEMPO}}
\newcommand{\PULSARMINER}{\texttt{PULSAR\_MINER}}
\newcommand{\PRESTOONGPU}{\texttt{PRESTO-ON-GPU}}

\newcommand{\dmunit}{pc\,cm$^{-3}$}
\newcommand{\pmunit}{mas\,yr$^{-1}$}
\newcommand{\msun}{{\rm M}_{\sun}}

\newcommand{\Pdot}{\dot{P}}
\newcommand{\Pbdot}{\dot{P}_{\rm b}}
\newcommand{\xdot}{\dot{x}}
\newcommand{\Pbdotobs}{\dot{P_{\rm b}}_{\rm ,obs}}
\newcommand{\Pdotobs}{\dot{P}_{\rm obs}}
\newcommand{\ra}{\alpha}
\newcommand{\dec}{\delta}

\newcommand{\Pb}{P_{\rm b}}
\newcommand{\Mp}{M_{\rm p}}
\newcommand{\Mc}{M_{\rm c}}
\newcommand{\Mtot}{M_{\rm tot}}
\newcommand{\omegadot}{\dot{\omega}}

\newcommand{\BWeff}{{\rm BW}_{\rm eff}}
\newcommand{\h}{^{\rm h}}
\newcommand{\m}{^{\rm m}}

\begin{document}

   \title{Discovery and timing of ten new millisecond~pulsars in the globular~cluster Terzan~5}

   \subtitle{}

   \author{
P.~V.~Padmanabh\inst{\ref{1},\ref{2}, \ref{3}}\thanks{E-mail: prajwal.voraganti.padmanabh@aei.mpg.de}
\and S.~M.~Ransom\inst{\ref{4}}\thanks{E-mail: sransom@nrao.edu}
\and P.~C.~C.~Freire\inst{\ref{3}}\thanks{E-mail: pfreire@mpifr-bonn.mpg.de}
\and A.~Ridolfi\inst{\ref{5},\ref{3}}
\and J.~D.~Taylor\inst{\ref{6}, \ref{7}}
\and C.~Choza\inst{\ref{4},\ref{8}}
\and C.~J.~Clark\inst{\ref{1},\ref{2}}
\and F.~Abbate \inst{\ref{5},\ref{3}}
\and M.~Bailes \inst{\ref{9}}
\and E.~D.~Barr \inst{\ref{3}}
\and S.~Buchner \inst{\ref{10}}
\and M.~Burgay\inst{\ref{5}}
\and M.~E.~DeCesar\inst{\ref{11}}
\and W.~Chen\inst{\ref{3}}
\and A.~Corongiu\inst{\ref{5}}
\and D.~J.~Champion\inst{\ref{3}}
\and A.~Dutta\inst{\ref{3}}
\and M.~Geyer\inst{\ref{12},\ref{10}}
\and J.~W.~T.~Hessels\inst{\ref{13},\ref{14}}
\and M.~Kramer\inst{\ref{3},\ref{15}}
\and A.~Possenti\inst{\ref{5}}
\and I.~H.~Stairs\inst{\ref{16}}
\and B.~W.~Stappers\inst{\ref{15}}
\and V.~Venkatraman~Krishnan\inst{\ref{3}}
\and L.~Vleeschower\inst{\ref{15}}
\and L.~Zhang\inst{\ref{17},\ref{9}}
          }
   \institute{
Max Planck Institute for Gravitational Physics (Albert Einstein Institute), D-30167 Hannover, Germany \label{1}
\and
Leibniz Universit\"{a}t Hannover, D-30167 Hannover, Germany \label{2}
\and
Max-Planck-Institut f\"{u}r Radioastronomie, Auf dem H\"{u}gel 69, D-53121 Bonn, Germany\label{3}
\and
National Radio Astronomy Observatory, 520 Edgemont Rd., Charlottesville, VA 22903, USA\label{4}
\and
INAF -- Osservatorio Astronomico di Cagliari, Via della Scienza 5, I-09047 Selargius (CA), Italy\label{5}
\and
Department of Physics and Astronomy, West Virginia University, P.O. Box 6315, Morgantown, WV 26506, USA\label{6}
\and
Center for Gravitational Waves and Cosmology, West Virginia University, Chestnut Ridge Research Building, Morgantown, WV 26505, USA\label{7}
\and
SETI Institute, 339 Bernardo Ave, Suite 200, Mountain View, CA 94043, USA\label{8}
\and
Centre for Astrophysics and Supercomputing, Swinburne University of Technology, P.O. Box 218, Hawthorn, VIC 3122, Australia\label{9} 
\and
South African Radio Astronomy Observatory, 2 Fir Street, Black River Park, Observatory 7925, South Africa\label{10}
\and
George Mason University, Fairfax, VA 22030, USA\label{11}
\and 
High Energy Physics, Cosmology and  Astrophysics Theory (HEPCAT), Department of Mathematics and Applied Mathematics, University of Cape Town,  Rondebosch 7701, South Africa\label{12}
\and
Anton Pannekoek Institute for Astronomy, University of Amsterdam,
Science Park 904, 1098 XH, Amsterdam, The Netherlands\label{13}
\and
ASTRON, Netherlands Institute for Radio Astronomy, Oude
Hoogeveensedijk 4, 7991 PD Dwingeloo, The Netherlands\label{14}
\and
Jodrell Bank Centre for Astrophysics, Department of Physics and Astronomy, The University of Manchester, Manchester M13 9PL, UK\label{15}
\and
Dept. of Physics and Astronomy, UBC, 6224 Agricultural Road, Vancouver, BC V6T 1Z1 Canada\label{16}
\and
National Astronomical Observatories, Chinese Academy of Sciences, A20 Datun Road, Chaoyang District, Beijing 100101, China\label{17}
            }
   \date{}

   \abstract{ We report the discovery of ten new pulsars in the globular cluster Terzan 5 as part of the Transients and Pulsars with MeerKAT (TRAPUM) Large Survey Project. We observed Terzan 5 at L-band (856--1712 MHz) with the MeerKAT radio telescope for four hours on two epochs, and performed acceleration searches of 45 out of 288 tied-array beams covering the core of the cluster.  We obtained phase-connected timing solutions for all ten discoveries, covering nearly two decades of archival observations from the Green Bank Telescope for all but one. Highlights include  PSR~J1748$-$2446ao which is an eccentric ($e = 0.32$) wide-orbit (orbital period $\Pb = 57.55$ d) system. We were able to measure the rate of advance of periastron ($\omegadot$) for this system allowing us to determine a total mass of $3.17 \pm \, 0.02\, \msun$. With a minimum companion mass ($\Mc$) of $\sim 0.8 \, \msun$, PSR~J1748$-$2446ao is a candidate double neutron star (DNS) system. If confirmed to be a DNS, it would be the fastest spinning pulsar ($P = 2.27$ ms) and the longest orbital period measured for any known DNS system. PSR~J1748$-$2446ap has the second highest eccentricity for any recycled pulsar ($e \sim 0.905$)  and for this system we can measure the total mass ($1.997 \pm 0.006 \msun$) and estimate the pulsar and companion masses, ($1.700^{+0.015}_{-0.045} \msun$ and $0.294^{+0.046}_{-0.014} \msun$ respectively). PSR~J1748$-$2446ar is an eclipsing redback (minimum $\Mc \sim 0.34\, \msun$) system whose properties confirm it to be the counterpart to a previously published source identified in radio and X-ray imaging. We were also able to detect $\omegadot$ for PSR~J1748$-$2446au leading to a total mass estimate of  $1.82 \pm \, 0.07\, \msun$ and indicating that the system is likely the result of Case A Roche lobe overflow. With these discoveries, the total number of confirmed pulsars in Terzan 5 is  49, the highest for any globular cluster so far. These discoveries further enhance the rich set of pulsars known in Terzan 5 and provide scope for a deeper understanding of binary stellar evolution, cluster dynamics and ensemble population studies.     
   }


   \maketitle

\section{Introduction}
\label{sec:intro}

Radio pulsar searches in globular clusters (GCs) have yielded 325 discoveries in 42 different GCs\footnote{An up to date list can be found at \url{https://www3.mpifr-bonn.mpg.de/staff/pfreire/GCpsr.html}}, demonstrating that GCs are unusually efficient pulsar factories \cite[see e.g.][]{Camilo_Rasio_2005,Ransom_2008,Freire_2013}. The high core densities in GCs (up to $\sim$ $10^6$ stars $\rm pc^{-3}$) encourage the formation of binary systems and exchange interactions between binaries. In many of these exchanges, those neutron stars (NSs) that have crossed the "death line" (that is their radio emission turned off and hence have become undetectable), become members of binary systems. The evolution of their companions leads to mass transfer onto these NSs, spinning them up to millisecond spin periods. During this stage, these systems are observed as low mass X-ray binaries (LMXBs). The latter are $\sim 10^3$ times more abundant per unit stellar mass in GCs compared to the Galaxy \citep{Clark_1975}. Once accretion stops, these NSs  become detectable as radio millisecond pulsars \citep[MSPs; see e.g.][and references therein]{Alpar_1982, Radhakrishnan_Srinivasan_1982,Ivanova_2013}.

The remarkable rotational stability of MSPs ($\Pdot \sim 10^{-20}\, \rm s s^{-1}$) has allowed for precise measurements of astrometric, spin and binary parameters of various  systems in GCs. For MSPs in compact binary systems, when this precision is coupled with long timing baselines (> 10 years), it enables the measurement of general relativistic effects via measurement of Post-Keplerian (PK) parameters like the rate of advance of periastron ($\omegadot$), the Shapiro delay and the Einstein delay, which have allowed measurements of a few NS masses \citep[e.g.][]{Lynch_2012_NGC6544B, Ridolfi_2019_NGC1851A, Corongiu_2023_NGC6752A} and in one case tests of gravity theories \citep{Jacoby_2006_M15C}. Apart from individual systems, an ensemble of precisely timed pulsars  can help constrain the structural properties of GCs and place upper limits on the mass of a potential intermediate mass black hole in their centres \citep[e.g.][]{Prager_2017,Perera_2017,Freire_2017,Abbate_2018,Abbate_2019}. Furthermore, these pulsars can also probe the ionised gas in the intra-cluster medium \citep{Freire_2001a,Abbate_2018} and also along the line-of-sight \citep{Martsen_2022}.

The high core densities in GCs also promote the formation of a wide range of unique binary pulsars whose properties stand out from the pulsars typically found in the Galactic field. For example, most known binary pulsars in the Galactic field with substantial eccentricity are double neutron star systems where a supernova explosion of the companion induces an eccentricity \citep{Tauris_2017}. However, binary pulsars in GCs can undergo multiple close encounters with neighbouring stars inducing a significant eccentricity in these objects \citep{Phinney_1992,Heggie_Rasio_1996}. These encounters can further lead to an exchange of pulsar companions where usually a higher mass object replaces the lighter companion that spun up the pulsar \citep[e.g.][]{Verbunt_Freire_2014}, also termed a secondary exchange encounter. This is evidenced by the GCs with the largest rate of such encounters having several highly eccentric systems with fast spinning pulsars and a companion mass larger than expected from standard recycling scenarios \citep[e.g.][]{DeCesar_2015,Ridolfi_2019_NGC1851A}, which are therefore very likely to be secondary exchange products. Apart from the possibility of finding exotic and unique pulsars, the enhanced sensitivity of current generation telescopes like MeerKAT and FAST, provides further motivation for continuing GC pulsar searches. One such globular cluster rich with pulsars is Terzan 5.

Terzan 5 (Ter5 hereafter) has historically been one of the most widely studied GCs  and has yielded remarkable results across multiple wavelengths. Located in the inner bulge of the Galaxy at a distance of $D = 6.62 \pm 0.15 \, \rm kpc$ from Earth \citep{Baumgardt_Vasiliev_2021}, it is believed to be a remnant of primordial structures that are integral to Galaxy formation \citep{Ferraro_2009}.  It is massive \citep[$(1.09 \pm 0.08) \times 10^{6}\, \msun$; ][]{Baumgardt_Vasiliev_2021} and is among the GCs with the highest stellar encounter rate \citep[$\Gamma \sim$ 6800;][]{Bahramian_2013}. This is commensurate with Ter5 holding the record for hosting the largest number of confirmed MSPs in a GC (39; prior to this work) of which more than 50\% are in binary systems (20; prior to this work). In contrast, NGC 104 (also known as 47 Tuc) which hosts the second highest number of confirmed MSPs, has a stellar encounter rate nearly a factor of seven lower than Ter5 \citep[$\Gamma \sim$ 1000;][]{Bahramian_2013}. Owing to the excess reddening from the Galactic bulge at optical and infra-red wavelengths, Ter5 has been observed and analysed more in the high-energy end of the spectrum. Apart from  evidence for gamma-ray emission in GeV \citep{Abdo_2010} and TeV ranges \citep{HESS_2011}, at least 50 sources have been identified in X-rays \citep{Heinke_2006}. Recent deep X-ray studies have established cross-matches to several known pulsars \citep[see][and references therein]{Bogdanov_2021}. Conversely, sources with significant X-ray variability with no known pulsar counterparts have been attributed as potential candidates for `spider' type pulsars \citep[e.g.][]{Urquhart_2020}. These are pulsars where the companion material is being ablated away by the pulsar wind and this outflow material can obscure the pulsed radio emission leading to dynamic eclipses across an orbital period cycle. Depending on the companion mass, the "spider" systems can be classified as redbacks ($\Mc \sim 0.2-0.4 \msun$) or black-widow systems \citep[$\Mc << 0.1 \msun$;][]{Roberts_2013}.  The X-ray emission in these spider pulsars is usually caused by an intra-binary shock \citep[see e.g.][]{Merwe_2020}. These results have established the need for synergistic studies across the electromagnetic spectrum for an enhanced understanding not only of individual pulsars but also to trace the origins and formation of Ter5.

The radio pulsar searches in Ter5 have benefited from a combination of large telescopes combined with advanced instrumentation and unique search techniques. The first Ter5 discovery, PSR~J1748$-$2446A, was found using the Very Large Array using a standard periodicity search and is still the most compact binary known in Ter5 \citep[$P_{\rm b} = 0.07 \, \rm d$;][]{Lyne_1990}. Later, a Fourier domain acceleration search technique was developed which provides improved sensitivity to binary pulsars  \citep{Ransom_2002}. This technique, along with the S-band receiver and the SPIGOT pulsar backend \citep{Kaplan_2005} on the  Robert C. Byrd Green Bank Telescope (GBT) yielded a flurry of discoveries \citep{Ransom_2005}. Soon after, this setup also yielded the current record holder for the fastest spinning pulsar PSR~J1748$-$2446ad \cite[$P=1.39$ ms;][]{Hessels_2006}. More recent discoveries have benefited from the application of different search techniques. For example, application of an additional acceleration derivative or `jerk' dimension to the binary pulsar search space helped discover PSR~J1748$-$2446am \citep{Andersen_Ransom_2018}. PSR~J1748$-$2446ae was found using a dynamic power spectrum search technique where the orbital period is roughly of the order of the observation time span searched (Ransom et al., in prep). Three more pulsars were found via stacking of Fourier power spectra from hundreds of hours of archival GBT data \citep{Cadelano_2018}. 

The latest pulsar discovery Ter5an, benefited significantly from the enhanced sensitivity of the MeerKAT radio telescope despite using just 42 out of the 64 antennas \citep{Ridolfi_2021}. Apart from demonstrating the value of MeerKAT for pulsar surveys, this study demonstrated the benefits of using archival data from the GBT. Based on an initial orbital solution from MeerKAT, the archival data allowed the quick determination of a long baseline (16 yr) phase connected solution. This large baseline even allowed the detection of PK effects like  orbital period derivative ($\Pbdot$) and rate of advance of periastron ($\omegadot$), which yielded measurements of the systemic acceleration of the system in the potential of the cluster and its total mass ($\Mtot = 2.97 \, \pm \, 0.52 \, \msun$). Archival data has since proved valuable for timing several other MeerKAT GC pulsar discoveries \citep{Douglas_2022, Ridolfi_2022, Gautam_2022, Abbate_2022, Vleeschower_2022}.

Recently, \citet{Martsen_2022} published spectral indices and flux densities for most of the known pulsars in Ter5. The corresponding pseudo-luminosity showed a turnover in the `logN-logS' relation which describes the cumulative number of sources detectable at a given telescope sensitivity. This suggested that all the searches conducted so far are incomplete \citep[see Figure 3 in][]{Martsen_2022}. It also agrees with previous population synthesis simulations suggesting that more than 100 detectable pulsars (allowing for beaming) reside in Ter5 \citep{Bagchi_2011,Chennamangalam_2013}, thus motivating further searches.  

In this paper, we describe ten new millisecond pulsars discovered in Ter5 with the MeerKAT radio telescope. The paper is structured as follows. We present the observations with MeerKAT and GBT in Section \ref{sec:observations}. The corresponding data reduction and techniques applied for searching and timing the new pulsars are presented in Section \ref{sec:data_analysis}. We discuss the physical properties of individual discoveries in Section \ref{sec:discoveries_and_timing} and also provide phase connected timing solutions for nine of these pulsars. Section \ref{sec:masses} discusses the individual mass constraints of the pulsar and companion in some selected binary pulsar discoveries. In Section \ref{sec:discussion}, we provide a discussion on the scientific outcomes of the new discoveries individually as well as collectively and the prospects for future studies involving Ter5. Finally, we state our conclusions in Section \ref{sec:conclusions}.

\section{Observations}
\label{sec:observations}

\subsection{MeerKAT}
\label{subsec:mkt_observations}

\begin{figure*}
\centering
	\includegraphics[width=0.49\textwidth]{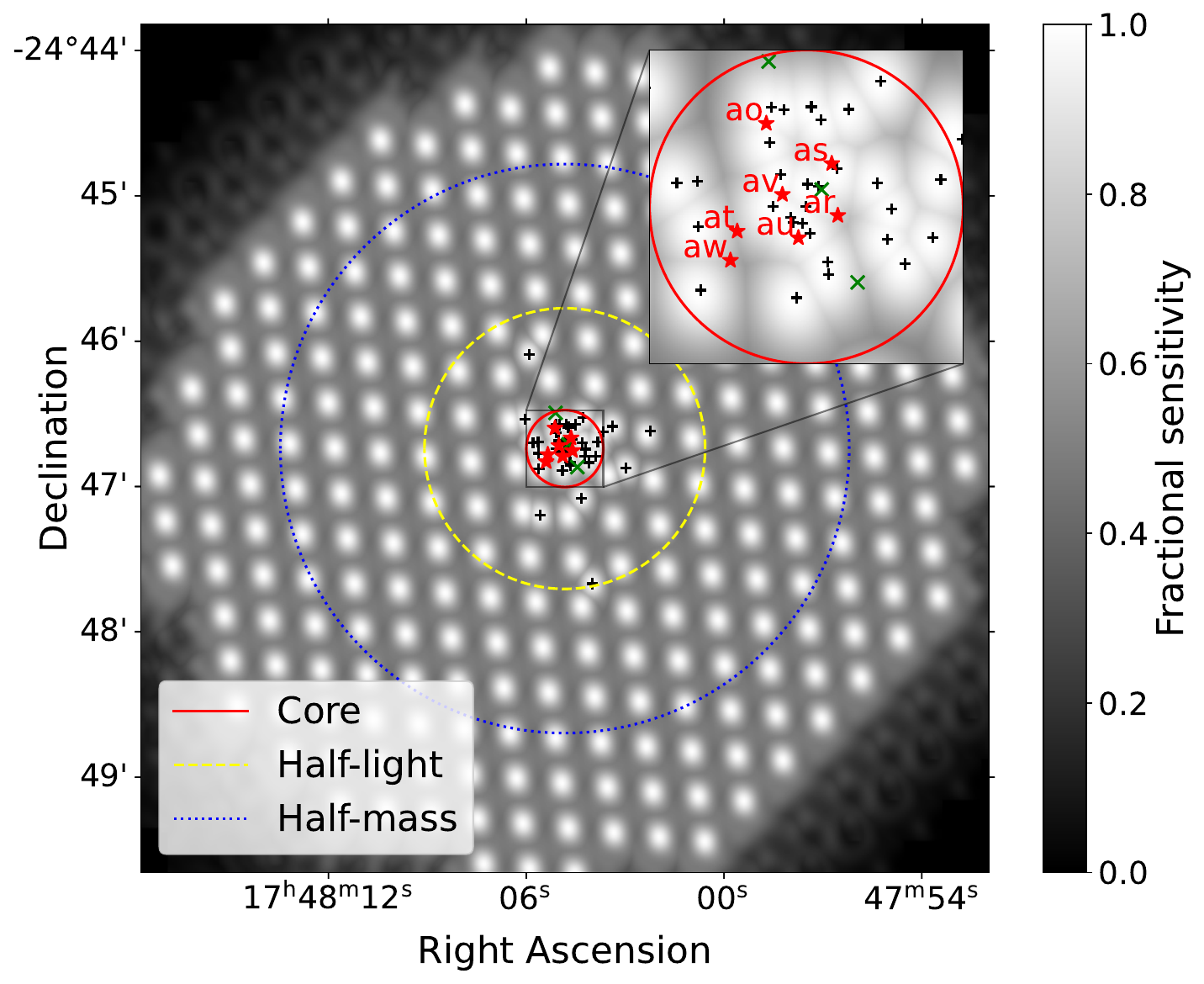}
	\includegraphics[width=0.49\textwidth]{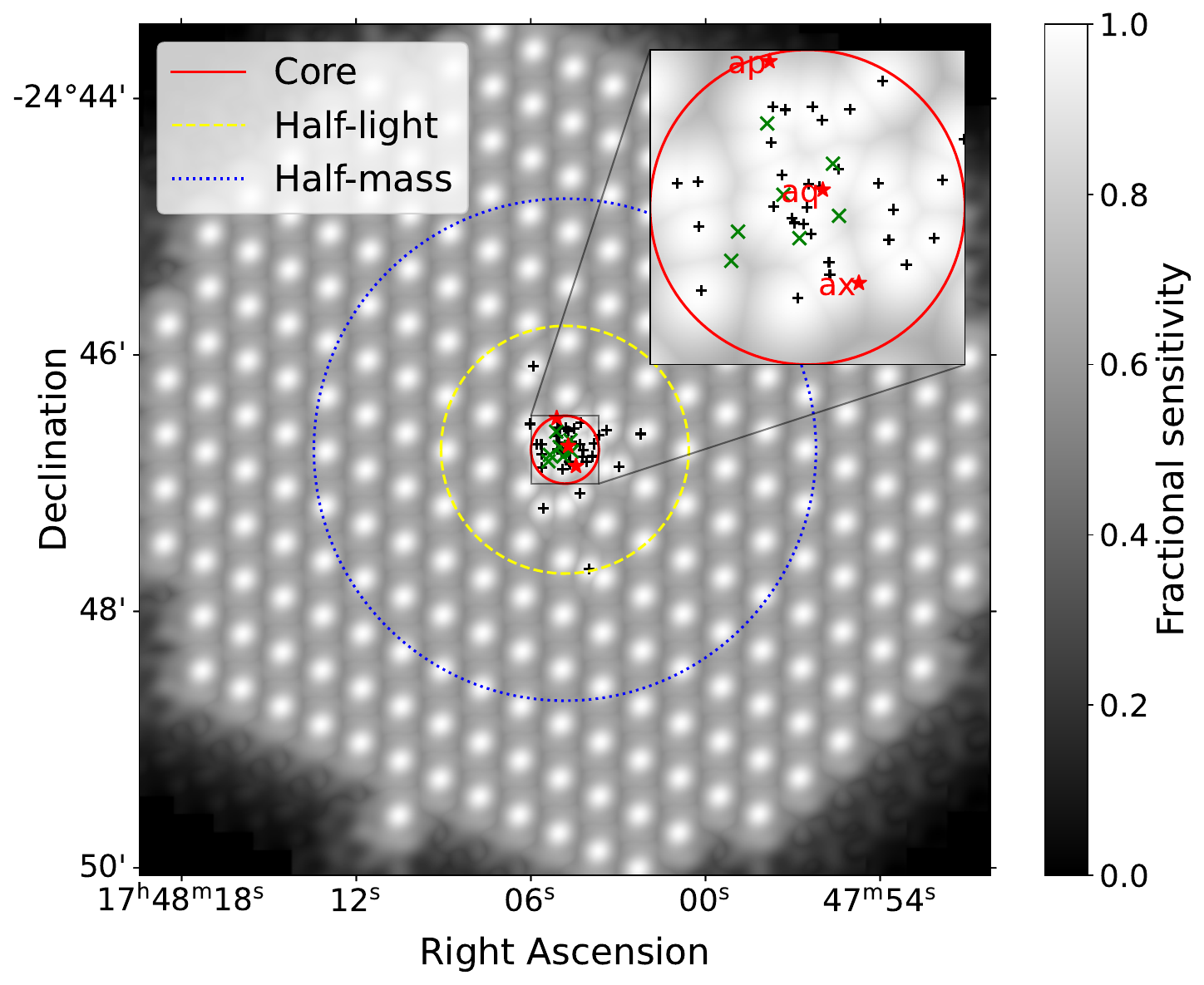}
	\caption{TRAPUM beam tilings of the two search observations (Epoch 1 and Epoch 2; see text) of Ter5, performed with MeerKAT on 05 Sep 2020 (left) and on 06 Jan 2021 (right) at L-Band. There are 288 beams in total per observation. Each individual beam has been overlaid with the corresponding point spread function. The vertical bar shows the reference gray scale for the fractional sensitivity with reference to the boresight position of the individual beam. The different concentric circles indicate the core radius (red), half-light radius (yellow-dashed) and half-mass radius (blue dotted). The known pulsars are marked in each tiling as black crosses. The discoveries made in each epoch are denoted as red stars along with their individual names. The green crosses indicate the positions of the discoveries but from the alternate epoch where the discovery was first made. The positions of the discoveries comes from timing.}
  	\label{fig:Ter5_tiling_patterns}
\end{figure*}

Ter5 was observed with the MeerKAT radio telescope using the L-Band (856--1712 MHz) receiver on 5 September 2020  (hereafter Epoch 1) and on 6 January 2021 (hereafter Epoch 2). This was done as part of the Globular Cluster pulsar survey within the umbrella of the Transients and Pulsars with MeerKAT (TRAPUM) large survey project \citep{Stappers_Kramer_2016}. Each observation spanned 4 hours and used 288 synthesised beams spanning roughly a 3\arcmin\, radius. Of the 288 beams, 38 were placed on positions of known pulsars\footnote{While there were 39 known pulsars previous to this work, Ter5al does not have a fully phase connected solution and hence a poor constraint on its position. This pulsar was thus not allocated an individual beam.}. The rest of the 250 beams were placed in a hexagonal packed tiling centered at the position of PSR~J1748$-$2446N (right ascension $\ra = 17\h48\m04\fs91$ and declination $\dec~=-24\degr46\arcmin53\farcs81$; Ransom et al., in prep) since this pulsar is located within the central core of Ter5.

The point spread function (PSF) of the individual beam and the corresponding tiling pattern was generated using the  \texttt{MOSAIC} software suite\footnote{\url{https://github.com/wchenastro/Mosaic}} \citep{Chen_2021} assuming an overlap factor of 70\%. This indicates that neighbouring beams shared a common boundary at a distance where the power level drops to 70\% from the centre of the PSF. Figure \ref{fig:Ter5_tiling_patterns} gives a detailed visualisation of the beam tiling pattern as well as the positions of known pulsars and the discoveries from this work overlaid across both observations.

Total intensity  \texttt{SIGPROC} format filterbank data for each individual synthesised beam were generated using Max-Planck-Institut f\"{u}r Radioastronomie's (MPIfR's) Filterbanking Beamformer User Supplied Equipment (FBFUSE) and was written to disk on MPIfR's Accelerated Pulsar Search User Supplied Equipment (APSUSE) high performance compute cluster. A detailed description of FBFUSE and APSUSE can be found in \citet{Padmanabh_2023} and references therein. The filterbank data were recorded at a sampling time of 76 $\mu$s and 4096 frequency channels with a channel bandwidth of  0.208 MHz. Additionally, we also recorded a single beam via the Pulsar Timing User Supplied Equipment (PTUSE) backend \citep{Bailes_2020} placed on the position of PSR~J1748$-$2446N in Epoch 1 and 2. The PTUSE data were recorded with a 9.57 $\mu$s time resolution with 1024 channels spanning 856 MHz after applying coherent dedispersion at  at a dispersion measure (DM) of 237 \dmunit, which is roughly the average of the DMs of all the previously known pulsars. Due to the limited capacity for long-term offline storage, the PTUSE data were later decimated in time from 9.57 $\mu$s to 75.29 $\mu$s and to 512 frequency
channels. This ensured that despite the reduction in size, the data fidelity was minimally affected (since the average dispersion measure for the pulsars in Ter5 is known, and the decimation in frequency accounts for that).

These observations offered multiple advantages compared to previous Ter5 observations conducted with MeerKAT as described in \citet{Ridolfi_2021}. Firstly, both observations were done with 56 antennas each thus providing a 33\% boost in telescope gain compared to the observations made by \citet{Ridolfi_2021}. Secondly, the semi-major axis of an individual tied-array beam  is $\sim$ 15 arc-seconds compared to 0.5 arcmin from before, owing to the usage of outer antennas that in turn increased the maximum baseline of the MeerKAT interferometer. Thus, any new discovery would already have a reasonable constraint on the  position. This, however, comes with the increased computational and data storage cost of having to process a large number of beams. Finally, the large number of beams tiled across a given patch of sky provides a nearly uniform sensitivity across a wide area compared to observations with single dish telescopes like Parkes or GBT where the sensitivity is reduced away from the boresight position.     

Apart from the above two observations, we also conducted five observations between 26 June 2023 and 7 July 2023 as part of a Shapiro delay campaign and to also derive orbital solutions for four of the ten discoveries made in Ter5 (further details are given later in Section \ref{sec:discoveries_and_timing}).  Finally, we also utilised archival PTUSE data recorded as part of the Globular cluster theme under the MeerTIME project \citep{Bailes_2020} between May 2019 and February 2020 (reported in \citealt{Ridolfi_2021}). Table \ref{tab:list_observations} summarises all the MeerKAT observations and the respective configuration parameters used.

\begin{table*}
\renewcommand{\arraystretch}{1.1}
\setlength{\tabcolsep}{0.2cm}
\footnotesize
\centering
\caption{List of the MeerKAT observations of Ter5 used for this work. $t_{\rm samp}$: sampling time; $N_{\rm pol}$: number of Stokes parameters; $f_{\rm c}$: central frequency; BW: observing bandwidth; $N_{\rm chan}$: number of frequency channels; $N_{\rm ant}$: number of antennas; $N_{\rm beam}$: number of tied-array beams. The `-orb' in the Observation id indicates observations made as part of campaigns to derive orbital solutions for newly discovered pulsars.}
\label{tab:list_observations}
\begin{tabular}{crrrrrcccccr}
\hline
Observation id            & \multicolumn{1}{c}{Observation date}    & \multicolumn{1}{c}{Start Time}       & \multicolumn{1}{c}{Length}    & \multicolumn{1}{c}{Primary Backend} & \multicolumn{1}{c}{$t_{\rm samp}$} & \multicolumn{1}{c}{$N_{\rm pol}$}& \multicolumn{1}{c}{$f_{\rm c}$}  & \multicolumn{1}{c}{BW}  & \multicolumn{1}{c}{$N_{\rm chan}$} & \multicolumn{1}{c}{$N_{\rm ant}$}  & $N_{\rm beam}$    \\
               & \multicolumn{1}{c}{}        & \multicolumn{1}{c}{(MJD)}            & \multicolumn{1}{c}{(s)}  & \multicolumn{1}{c}{}  & \multicolumn{1}{c}{($\upmu$s)} &  & (MHz)        & (MHz) &               &   \\
\hline
01L$^\dag$      & 27 May 2019 & 58630.813 & 9000 & PTUSE & 9.57 & 4 & 1284 & 642 & 768 & 42 & 1\\ 
02L-orb$^\dag$  & 26 Feb 2020 & 58905.048 & 12600 & PTUSE & 9.57 & 4 & 1284 & 856 & 4096 & 42 & 1\\ 
03L-orb$^\dag$  & 27 Feb 2020 & 58906.017 & 12600 & PTUSE & 9.57 & 4 & 1284 & 856 & 4096 & 42 & 1\\
04L-orb$^\dag$  & 28 Feb 2020 & 58907.022 & 12600 & PTUSE & 9.57 & 4 & 1284 & 856 & 4096 & 42 & 1\\
05L-orb$^\dag$  & 28 Feb 2020 & 58907.326 &  12600 & PTUSE & 9.57 & 4 & 1284 & 856 & 4096 & 42 & 1\\
06L$^{*1}$       & 05 Sep 2020 & 59097.671 & 14400 & APSUSE & 76.56 & 1 & 1284 & 856 & 4096 & 56   & 288 \\
07L$^{*2}$       & 06 Jan 2021 & 59342.278 &  14400 & APSUSE & 76.56 & 1 & 1284 & 856 &  4096 & 56   & 288 \\
08L-orb$^*$   & 26 Jun 2023 & 59342.586 &  14400 & APSUSE & 76.56 & 1 & 1284 & 856 &  4096 & 60   & 22 \\
09L-orb$^*$   & 27 Jun 2023 & 59355.253 &  14400 & APSUSE & 76.56 & 1 &  1284 & 856 &  4096 & 60   & 22 \\
10L-orb$^*$   & 28 Jun 2023 & 59355.628 &  14400 & APSUSE & 76.56 & 1 &  1284 & 856 &  4096 & 60   & 22 \\
11L-orb$^*$   & 30 Jun 2023 & 59358.242 &  14400 & APSUSE & 76.56 & 1 &  1284 & 856 &  4096 & 60   & 22 \\
12L-orb$^*$   & 07 July 2023 & 59360.607 &  14400 & APSUSE & 76.56 & 1 &  1284 & 856 &  4096 & 60   & 22 \\
\hline \hline
\multicolumn{8}{l}{$^*$ TRAPUM observations with single beam PTUSE data recorded in parallel.}\\
\multicolumn{8}{l}{$^\dag$ MeerTIME observations}\\
\multicolumn{4}{l}{$^{1}$ refers to Epoch1 and $^{2}$ to Epoch 2 respectively (see text)}\\
\end{tabular}\\
\end{table*}

\subsection{GBT}
\label{subssec:gbt_obs}

Apart from MeerKAT data, we also used archival data from GBT observations for deriving long-term phase connected solutions for all our discoveries. GBT data from three backends, namely, SPIGOT \citep{Kaplan_2005}, GUPPI \citep{DuPlain_2008}, and VEGAS \citep{Prestage_2015} were used. Pulsar/GC observations with these back-ends are described in more detail by \cite{Ransom_2005,Cadelano_2018} and \cite{Martsen_2022} respectively. This represents a total of $\sim$ 130 observations across a timeline spanning 2004$-$2021.

\section{Data analysis}
\label{sec:data_analysis}

\subsection{Cleaning and subbanding}
\label{subsec:data_clean_and_subband}

Before searching, the TRAPUM filterbank data from all beams of all epochs were put through a crucial preprocessing step. The data were first cleaned using the Inter-Quartile Range Mitigation algorithm \citep{Morello_2022}, thus significantly reducing the impact of radio frequency interference (RFI). Following this, groups of 16 channels were dedispersed at the nominal cluster DM of 237 \dmunit\, reducing the total number of channels to 256. This subbanding process  not only reduced data volume but also sped up subsequent processing steps. Post subbanding, the raw data recorded with 4096 channels were deleted to free up storage space on APSUSE.

\subsection{Search strategy}
\label{subsec:search_strategy}

We used \PULSARMINER\, \citep{Ridolfi_2021}, a Python-based wrapper for \PRESTO\, \citep{Ransom_2011},  for implementing acceleration searches on the subbanded filterbank data. \PULSARMINER \, has already proven to be successful in discovering multiple pulsars across various globular clusters \citep{Ridolfi_2021, Ridolfi_2022, Vleeschower_2022, Abbate_2022, Chen_2023}. First, we dedispersed the time series across 500 different DM trials between 230 and 250 \dmunit\, with a step size of 0.05 \dmunit. The upper and lower DM trial limits were chosen to extend the range slightly beyond the lowest (Ter5Q; DM = 234.50 \dmunit) and highest DM (Ter5D; DM= 243.83 \dmunit) among the known pulsars in Ter5. Our search used incoherent harmonic summing with 8 harmonics and a threshold of $z_{\max}\, =\, 200$ where $z_{\max}$ is the maximum spectral drift (in terms of frequency bins) due to linear acceleration. The limit  placed on the value of $z$ was to strike a balance between sensitivity to binary pulsars and the computational expense of running the searches. In general, $z$ depends on the duration of the observation ($T_{\rm obs}$) and the line-of-sight component of the orbital acceleration of the pulsar ($a_l$) as \citep[see e.g.][]{Andersen_Ransom_2018}
\begin{equation}
\label{eq:z_a}
z = \frac{T_{\rm obs}^2a_lh}{cP},
\end{equation}
where $P$ is the pulsar's spin period, $h$ is the harmonic number and $c$ denotes the speed of light. Using Equation \ref{eq:z_a}, we can derive the maximum acceleration $a_{l, \max}$ we are sensitive to based on the value of $z_{\max}$. For example, a pulsar spinning at P = 2 ms and $z = 200$ would experience $a_{l, \max}$ = 9.2 $\rm ms^{-2}$ for a 1 hr data segment when searching only the fundamental frequency (h=1).  

In order to be sensitive to compact binary pulsars, we searched non-overlapping data segments of 0.5 hr, 1 hr, 2 hr, and the full 4 hr time span. This segmented search approach enabled the best sensitivity to pulsars with an orbital period above 5 hours. We also applied a sifting algorithm to retain candidates which are detectable in at least 3 neighbouring DM trials and are above a 4 $\sigma$ threshold in Fourier significance. Candidates that made this cut were folded and visually inspected. These criteria led to an average of 1000 candidates per beam. The parameters of promising candidates were used to fold data from neighbouring beams from the same epoch and the closest beams in the alternate epoch. A detection in more than one epoch gave a strong indication that the pulsar candidate was real. 

The \PULSARMINER\,  pipeline using the above-mentioned configuration parameters was run on the ATLAS\footnote{\url{https://www.aei.mpg.de/atlas}} supercomputer operated by the Max Planck Institute for Gravitational Physics in Hanover, Germany. In order to speed up processing, \PULSARMINER\, was run with \PRESTOONGPU \footnote{\url{https://github.com/jintaoluo/presto_on_gpu}} enabling the acceleration search routine to be run on GPUs and speed up this step by a factor of 20-30.

\subsection{Deriving orbital solutions} 
\label{subsec:orbital_solution}

Discoveries that showed a non-zero acceleration or a change in barycentric period across epochs strongly indicated  that the pulsars were in binary systems. As a first step, we plotted the observed spin period and acceleration parameters from multiple epochs on a period-acceleration diagram \citep{Freire_2001b}. If all the points tracked out an ellipse, this gave an immediate indication that the pulsar is most likely in a near-circular orbit. We then fit a parabola (as explained in \citealt{Freire_2001b}) to derive the spin period of the pulsar $P$ and two binary parameters, the orbital period ($\Pb$) and the semi-major axis of the pulsar's orbit ($a$) projected along the line of sight and expressed in time units ($x \equiv a \sin i / c$, where $i$ is the generally unknown orbital inclination). This ephemeris was used in turn as an input to \texttt{fit\_circular\_orbit.py} from \PRESTO, which also fits for the time of passage through periastron ($T_0$) by assuming that a single sinusoidal modulation will fit all detections. If the initial spin period vs. acceleration pattern was not an ellipse, we used \texttt{fitorb.py} from \PRESTO\, to also fit for eccentricity ($e$) and longitude of periastron ($\omega$).

In order to obtain more data points for better constraining the orbital solution, we first used the initial orbital solution (obtained with relatively few epochs) to predict the expected spin period and acceleration in other epochs. We then folded data from other epochs using the predicted spin and acceleration values as starting input parameters. This method was successful in increasing the overall number of detections and led to a more robust orbital solution when put through another iteration of \texttt{fit\_circular\_orbit.py} or  \texttt{fitorb.py}. The routine to calculate the expected spin period and acceleration for a given epoch is now available as a Python routine termed \texttt{binary\_info.py} in \PRESTO.   

Depending on the segment length in which the discovery was made, a different strategy was applied to derive a rough orbital solution. For example, if discoveries were made in the shortest data segments of 30 min, we also attempted to get detections in other 30-min data segments  across the 4-hr observation span by restricting the input DM, spin and spin derivative parameters to search. Multiple closely spaced detections helped to easily break the degeneracy in obtaining a unique orbital solution. However, if the discovery was made in the full 4-hr data span, the pulsar was searched with a refined search space but in archival GBT data. If enough detections could be obtained (particularly, containing closely spaced detections spanning up to a day), we could be confident of fitting for an orbital solution.  For those pulsars where we could not obtain enough detections with GBT data, we used the five follow-up observations spanning 10 days (explained in Section \ref{subsec:mkt_observations} earlier) as a means to obtain closely spaced detections and ease the orbit solving process.  

\subsection{Timing}
\label{subsec:timing}

In order to obtain a precise estimate of the astrometric, spin and binary parameters, we attempted to extract times-of-arrival (TOAs) from all available data (from MeerKAT and GBT). Firstly, we folded the data using \texttt{prepfold} from \PRESTO\ with the best ephemeris derived from the orbital solution described earlier. For isolated pulsars, the best barycentric spin period and DM served as starting points to create an ephemeris to build on. Folding the data quickly revealed more detections across several epochs from which more TOAs could be extracted and in turn help in obtaining long-term phase connected solutions. Topocentric TOAs were extracted from the folded \PRESTO\, archives files (\textit{pfd} format) using \texttt{get\_TOAs.py} after cross-correlating an analytical template to all detected pulse profiles. This analytical template was obtained using \texttt{pygaussfit.py} from \PRESTO\, by fitting multiple Gaussian profiles with varying centroid positions, widths and heights. Depending on the brightness of each detection, the number of TOAs extracted varied from observation to observation. 

Finally, all the TOAs were fit for a timing model using \TEMPO\, \citep{Nice_2015}. We applied a scaling factor (known as EFAC) for each of the different subsets of TOAs derived from different observing back-ends. This factor increases the uncertainty values of the TOAs and in turn  makes the overall reduced $\chi^2$ value close to one. Some discoveries had a large number of bright detections which allowed for manually phase connecting the pulsar across the entire timing baseline. However, there were some pulsars where manual attempts failed. In these cases, we applied an automated timing procedure termed as the Algorithmic Pulsar Timer for Binaries\footnote{\url{https://github.com/Jackson-D-Taylor/APT}} \citep[APTB;][]{Taylor_2024}  that could help determine the exact number of rotation counts between the sparsely sampled data points. APTB expands on the Algorithmic Pulsar Timer \citep{Phillips_2022} algorithm by implementing techniques required for phase connecting binary systems in an automated manner. Furthermore, APTB uses robust statistical tests including the dependence on the reduced  $\chi^2$ (similar to \texttt{DRACULA}, developed by \citealt{Freire_Ridolfi_2018}) for decision making while building phase connection between data points. APTB also adds new parameters to fit on-the-fly, depending on lengths of data that are already phase connected from previous steps. A detailed description of this automated routine is described in \citealt{Taylor_2024}. One pulsar (Ter5ap; see Section \ref{subsec:Ter5ap}) could not be solved by APTB (or \texttt{DRACULA}), but was eventually solved by a new method currently in development (Clark \& van Haasteren, in prep), which greatly speeds up the phase-connection procedure by exploiting covariances between the uncertain rotation counts between observations to avoid costly model re-fitting. 

To describe the orbits, we have used several orbital models available in \TEMPO\, based on the theory-independent Damour \& Deruelle ("DD") orbital model \citep{Damour_Deruelle_1986}, which is used to estimate the Keplerian and PK parameters of the pulsar's orbit described in the next subsection. The first, known as "DDFWHE", is nearly identical to the DD model, except that it uses the orthometric parameterisation of the Shapiro delay \citep{Freire_Wex_2010}. A second variation of the DD model, known as "ELL1", is especially suited for very low-eccentricity orbits \citep{Lange_2001}. A third model, known as "BTX" \citep{Shaifullah_2016}, is derived from the Blandford-Teukolsky ("BT") model \citep{Blandford_Teukolsky_1976} and allows a non-predictive description of the random orbital period variations observed for several eclipsing systems.
The special "DDGR" model \citep{Damour_Deruelle_1986}  assumes the validity of general relativity (GR) to estimate masses directly and self-consistently from all observed relativistic effects. Details on these models and their parameters are given later in Section \ref{sec:discoveries_and_timing} as and when they are used to describe different pulsars.

\subsection{Derived parameters from Keplerian and Post-Keplerian parameter measurements}
\label{subsec:post_keplerian}

While fitting for different timing models with the measured TOAs, we also derived constraints on the properties of some of the discovered binaries based on estimates from Keplerian and PK parameters. The Keplerian parameters correspond to the five aforementioned orbital parameters, namely: $\Pb$, $x$, $e$, $\omega$, and $T_0$. When dealing with pulsars in compact binary systems, higher order relativistic effects become more prominent in the timing analysis, which can be quantified by additional PK parameters in a theory-independent way. These PK parameters depend on the Keplerian orbital parameters and the individual masses as prescribed by a specific gravity theory. For all cases below, we make the assumption that GR describes the strong-field gravity regime, unless stated otherwise. We give details of the derived and Keplerian and PK parameters below.

\textit{(i) Mass function}: From Kepler's third law, we can derive the mass function $f$ of the system from two of the Keplerian parameters of the pulsar's orbit
\begin{equation}
\label{eqn:kepler_third_law}
    f = \frac{(\Mc \sin i)^3}{(\Mp + \Mc)^2} = \frac{4 \pi^2}{T_{\odot}} \frac{x^3}{\Pb^2},
\end{equation}
where $i$ is the orbital inclination $\Mp$ is the pulsar mass and $\Mc$ is the companion mass divided by a solar mass ($\msun$) and $T_{\odot} = {\cal G M}^{\rm N}_{\odot} / c^3 =  4.925490947...\, \rm \upmu s$ is an exact constant, the solar mass parameter (${\cal G M}^{\rm N}_{\odot} \simeq G \msun$, \citealt{Prsa_2016}, where $G$ is Newton's gravitational constant) in time units. This high precision is warranted by the fact that the product $G \msun$ is known much more precisely than either $G$ or $\msun$.

Owing to the $\sin i \leq 1$ limit, one can derive a lower limit for $\Mc$ assuming a certain fixed value for $\Mp$. In this paper, we assume $\Mp = 1.35\, \msun$ for deriving the minimum companion mass unless stated otherwise.

\textit{(ii) Rate of advance of periastron} ($\omegadot$): This PK parameter quantifies the rate of change in the longitude of the periastron, which determines the orientation of the orbit with respect to the observer's line of sight. If the effect is purely relativistic, then the effect is always positive. Assuming the validity of GR we can express $\omegadot$ as a function of the total mass of the system  ($\Mtot = \Mp + \Mc$) as \citep[e.g.][]{Taylor_Weisberg_1982}:
\begin{equation}
\label{eqn:mtot_from_omegadot}
    \omegadot = 3 \left(T_{\odot} \Mtot \right)^{2/3}\, n_{\rm b}^{5/3}\, \frac{1}{1 - e^2},
\end{equation}
where $n_{\rm b} = 2 \pi / P_{\rm b}$.

\textit{(iii) Shapiro delay}: In cases where the orbit of a binary pulsar is viewed nearly edge-on, there is also a possibility of measuring a time delay in the pulsar signal owing to the gravitational influence of the companion. This is known as the Shapiro delay \citep{Shapiro_1964}. There are several ways to parameterise it. When there is a weak detection of the Shapiro delay, especially if the orbital inclination is not very close to $90\deg$, then the use of the "orthometric" parameters \citep{Freire_Wex_2010} avoids strong correlations. Assuming the validity of GR, the orthometric ratio ($\varsigma$) and amplitude ($h_3$) are given by:
\begin{eqnarray}
\varsigma & = & \frac{\sin i}{|\cos i| + 1} \\
h_3 & = & \Mc T_{\odot} \varsigma^3.
\end{eqnarray}
Even a faint detection of this effect can, when combined with a measurement of $\omegadot$, result in very precise mass measurements (e.g., \citealt{Martinez_2015,Stovall_2019,Mckee_2020}). As described later in Section \ref{sec:masses}, the same technique has been applied to derive precise mass measurements for three binaries in Ter5.

\textit{(iv) Variation of the orbital period} ($\Pbdot$): In some systems, this variation is caused by the emission of gravitational waves, which provide a precise test of gravity theories (e.g., \citealt{Taylor_Weisberg_1982,Taylor_Weisberg_1989,Kramer_2021} and references therein). However, for the binaries described in this paper, this effect is very small compared to the effect on $\Pbdotobs$ caused by the change in the Doppler factor of the binary system due to its motion. Differentiating the expression for the Doppler factor as a function of time, and assuming a negligible intrinsic variation of the orbital period, we obtain \citep{Damour_Taylor_1991}:
\begin{equation}
\label{eqn:pbdot_measured}
   \left(\frac{\Pbdot}{\Pb}\right)_{\rm obs} =  \frac{a_{\rm cluster}}{c} + \frac{a_{\rm gal}}{c}   + \frac{\mu^2 D}{c},
\end{equation}
where $a_{\rm cluster}$ represents the acceleration contribution from the pulsar in the gravitational potential of the cluster, $a_{\rm gal}$ is the contribution from the Galactic field potential and the last term is the contribution from the Shklovskii effect \citep{Shklovskii_1970} owing to transverse motion of the pulsar. This term depends on the total proper motion
$\mu = \sqrt{\mu_{\alpha}^2 + \mu_{\delta}^2}$ (where $\mu_{\alpha}$ is the proper motion contribution along right ascension and  $\mu_{\delta}$ along declination) as well as the distance of Ter5 from the Earth $D$. Of these effects, $a_{\rm cluster}$ is usually dominant. Thus, by estimating $a_{\rm gal}$ and $a_{\rm shk}$, we can determine $a_{\rm cluster}$ from precise measurements of $\Pbdotobs$. Those estimates can be used to constrain the mass model of the cluster \citep[see e.g.][]{Prager_2017}.

In addition, it allows us to uniquely solve for the true value of the intrinsic spin period derivative of the pulsar ($\dot{P}_{\rm true}$) as:
\begin{equation}
\label{eqn:p1_measured}
   \left(\frac{\Pdot}{P}\right)_{\rm obs} = \left(\frac{\Pdot}{P}\right)_{\rm true} + \frac{a_{\rm cluster}}{c} + \frac{a_{\rm gal}}{c}   + \frac{a_{\rm shk}}{c},
\end{equation}
Subtracting Equation \ref{eqn:pbdot_measured} from \ref{eqn:p1_measured} means we can estimate $\dot{P}_{\rm true}$ as 
\begin{equation}
\label{eqn:p1_pbdot_relation}
\dot{P}_{\rm true} = \Pdotobs -  \frac{\Pbdotobs}{\Pb}\, P.
\end{equation}
Once $\dot{P}_{\rm true}$ is obtained, other derived parameters like the characteristic surface magnetic field ($B = 3.2\times10^{-19}P \dot{P}_{\rm true}$) and characteristic age ($\tau_c = P / (2 \dot{P}_{\rm true})$) can also be obtained. 

\textit{(v) Variation of the projected semi-major axis} ($\xdot$): A change in the projected semi-major axis can be a consequence of changes in the physical size of the orbit or a change in $i$ from the changing viewing angle due to the motion of the binary. There are multiple factors that contribute to it including PK effects \citep[see e.g.][]{Lorimer_Kramer_2004}. Here we assume that the contribution from effects like aberration, Doppler modulation, gravitational wave damping and spin orbit coupling are negligible and the primary contribution is from proper motion. The observed $\xdot$ can be written as \citep[see][]{Arzoumanian_1996,Kopeikin_1996}:     
\begin{equation}
\label{eqn:xdot_contribution}    
\left(\frac{\dot{x}}{x}\right)_{\rm obs} = \left(\frac{\dot{x}}{x}\right)_\mu = 1.54 \times 10^{-16}\, \cot\, i\, (-\mu_{\alpha} \sin\Omega + \mu_{\delta} \cos\Omega),   
\end{equation}
where $\mu_{\alpha}$ and $\mu_{\delta}$ are the proper motion terms in right ascension and declination (expressed, in this equation, in \pmunit) respectively; $i$ is the orbital inclination and $\Omega$ is the longitude of ascending node. We can place an upper limit on the maximum contribution from proper motion and constrain $i$:
\begin{equation}
\label{eqn:i_constraint_from_xdot}  
\tan i < 1.54 \times 10^{-16}\, \mu \left(\frac{x}{\dot{x}}\right)_\mu,
\end{equation}
where $\mu$ is also expressed in \pmunit.
Placing this limit on the inclination angle can in turn place limits on the individual masses of the system. 

\textit{(vi) Einstein delay} ($\gamma$): This PK parameter quantifies a delay caused by the variation of the gravitational redshift and special relativistic time delay with orbital phase. In GR, it can be related to the binary masses as:
\begin{equation}
\label{eq:gamma}
\gamma = n_{\rm b}^{-\frac{1}{3}} T_{\odot}^{\frac{2}{3}} e \frac{M_{\rm c} (M_{\rm tot} + M_{\rm c}) }{M_{\rm tot}^{\frac{4}{3}}}.
\end{equation}
However, this effect is hard to measure for wide, eccentric binaries with a small change of $\omega$ within the timing baseline. This has only been achieved successfully for one wide binary, PSR~J0514$-$4002A (with $P_{\rm b} = 18.8 \rm d$, \citealt{Ridolfi_2019_NGC1851A}). This was done by showing first that the effect of $\gamma_{\rm E}$ on the timing of these systems is a linear variation of the projected semi-major axis of the pulsar's orbit ($\dot{x}_{\gamma}$, see their eq. 25). Therefore, this term superposes inevitably with the $\dot{x}$ from other causes, like the effect of the proper motion discussed above, $\dot{x}_\mu$. Only if the expected values of $\dot{x}_{\mu} << \dot{x}_{\gamma}$ can measure $\gamma$, as was done for PSR~J0514$-$4002A.

\subsection{Localisation}
\label{subsec:localisation}
The narrow width of the synthesised tied-array beams allows for an instantaneous localisation of new discoveries with an uncertainty of tens of arc-seconds (at L-Band). However, if a pulsar is detected in multiple neighbouring beams with different signal-to-noise (S/N), one can better constrain the position. The \texttt{SeeKAT} software \citep{Bezuidenhout_2023} implements such  
an algorithm  by measuring the likelihood of the true pulsar position weighted by the point spread function (PSF) and the corresponding different S/N values in neighbouring beam positions. This method can provide a sub-arcsecond localisation and has demonstrably eased the phase connection process when fitting for the position of the pulsar \citep[see][and references therein]{Bezuidenhout_2023}.  
As explained later in Section \ref{sec:discoveries_and_timing}, we used \texttt{SeeKAT} to localise two of the faint discoveries and this quickly led to a unique phase connected timing solution for one of these.

\subsection{Radiometer flux density estimate}
\label{subsec:flux_density}

In order to obtain the limiting flux density of the survey ($S_{\mathrm{min}}$), we used the radiometer equation \citep{Morello_2020} given as:
\begin{equation}
\label{eq:radiometer_equation}
S_{\mathrm{min}} = \frac{ {\mathrm {S/N}} \,\, \beta  \,\, (T_{\mathrm{sys}})} { G \epsilon \sqrt{n_{\mathrm{pol}}  \,\,  \BWeff \,\, t_{\mathrm{obs}}}}  \,\, \sqrt{\frac{\delta}{1 - \delta}},
\end{equation}
where $n_{\mathrm{pol}}$ is the number of polarisations which is 2. The S/N threshold was chosen to be 10. $\beta$ accounts for digitisation losses and was chosen as 1.0 due to very minimal loss (< 0.01\%) in 8-bit data \citep{Kouvenhoven_2001}. The gain of the telescope $G$ was chosen as 2.5 K/Jy, corresponding to 56 dishes of MeerKAT representative of most of our observations. The system temperature was chosen as 26 K after taking into account the receiver temperature and the sky temperature at L-Band. $t_{\mathrm{obs}}$ is the observation time which was chosen as 4 hours. About 25 \% of the band is unusable leading to the effective bandwidth  $\BWeff$ of 642 MHz. The duty cycle of the pulsar $\delta$ was chosen as 10\%. Finally, we assumed a search efficiency factor of $\epsilon$=0.7 based on the work by \citealt{Morello_2020} when accounting for FFT based searches up to 8 incoherent harmonic sums. Finally, this is the best case scenario assuming maximum sensitivity per synthesised beam. Given the overlap factor of 0.7 between beams, the worst case scenario is a limiting flux density of 16.78 $\mu$Jy.    

Additionally, we also used Equation \ref{eq:radiometer_equation} without $\epsilon$ to obtain flux density estimates and corresponding pseudo-luminosity values ($L = S_{\mathrm{min}}D^2$ where $D$ is the distance to Ter5) for all the discovered pulsars. These values are given in Tables \ref{tab:eccentric} to \ref{tab:other_pulsars}. It is important to note that the flux values are subject to significant changes owing to refractive scintillation as well as different spectral indices per pulsar \citep[e.g.][]{Martsen_2022}.

\subsection{X-ray source cross matching}
\label{subsec:x-ray_cross_matching}

\begin{figure*}
    \centering
    \includegraphics[width=\textwidth]{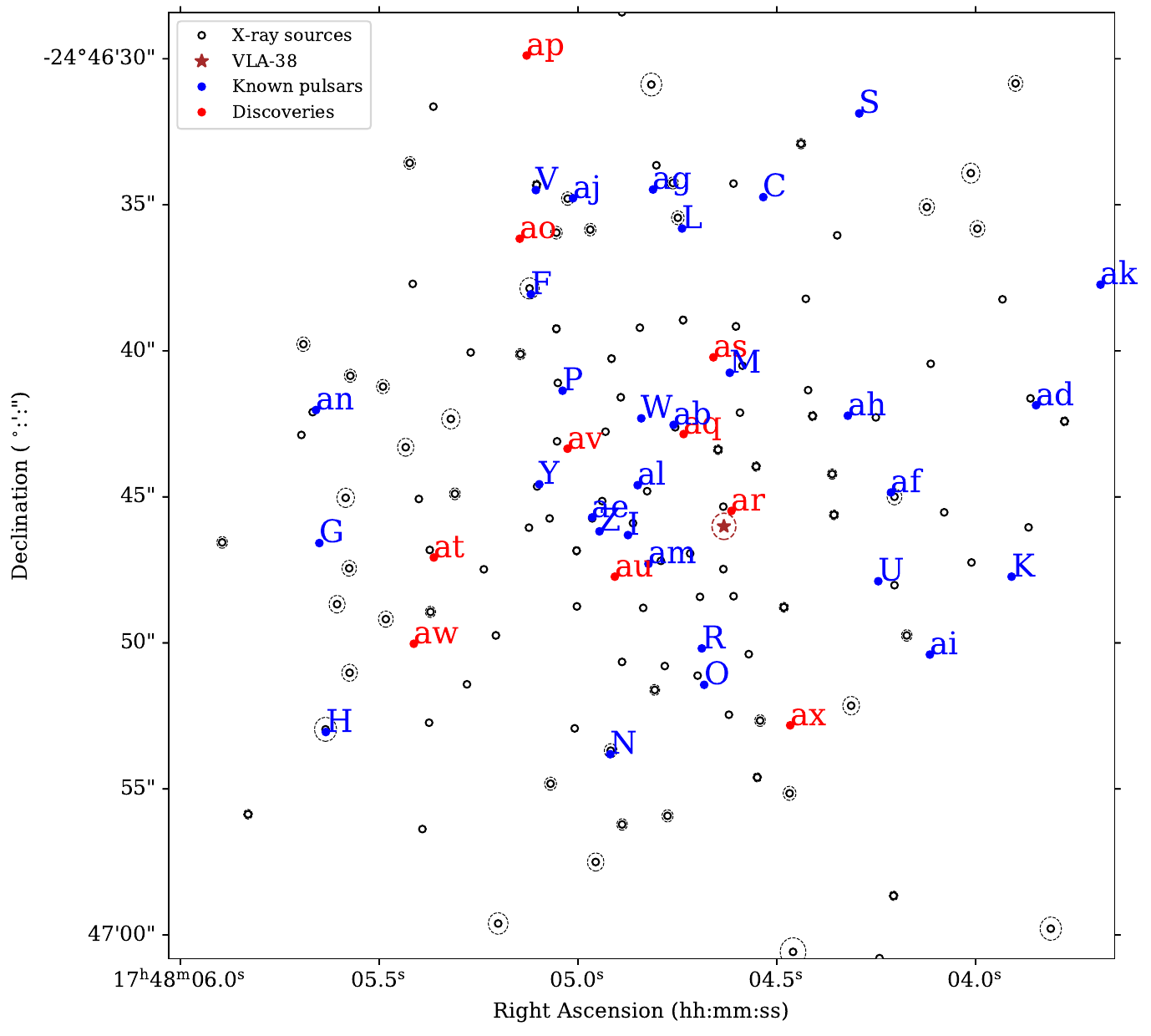}
    \caption{Ter5 field shown with each axis approximately spanning  1.5 times the core radius of the cluster. The positions of the X-ray sources (black rings), known pulsars (blue dots) as well new discoveries (red dots) are also overlaid. Particular emphasis is made on the VLA-38 source (brown star) obtained from radio imaging and whose coordinates are reported in \citealt{Urquhart_2020}. The timing position of Ter5ar (green star) is shown to overlap significantly with VLA-38 as well as with an X-ray source (CXOU J174804.63-244645.2) in the backdrop. The dashed lines surrounding the X-ray sources and VLA-38 are the 3-$\sigma$ positional uncertainties.}
    \label{fig:Ter5_xray}
\end{figure*}

As mentioned in Section \ref{sec:intro}, many  millisecond pulsars in Ter5 are known to have associations with X-ray sources and such a multi-wavelength study could help better understand the properties of each individual source. We used the catalogue provided by \citet{Bahramian_2020} of all X-ray sources in globular clusters to check for associations with any discoveries made. Figure \ref{fig:Ter5_xray} depicts the region within 1.5 times the core radius of the Ter5 field, with X-ray sources, known pulsars, and new pulsar discoveries plotted. We used a conservative simple cut of 0.5 arc-second as the maximum separation between the X-ray source and the discovered pulsars to suggest any association. This cut was chosen to mitigate the impact of any errors from previously applied source extraction methods. We comment on potential (not certain) cross-matches between the position  obtained from radio timing and those reported from X-ray imaging in \citealt{Bahramian_2020} in the next section.   

\section{Discoveries and their properties}
\label{sec:discoveries_and_timing}
\begin{figure*}
    \centering
    \includegraphics[width=\textwidth]{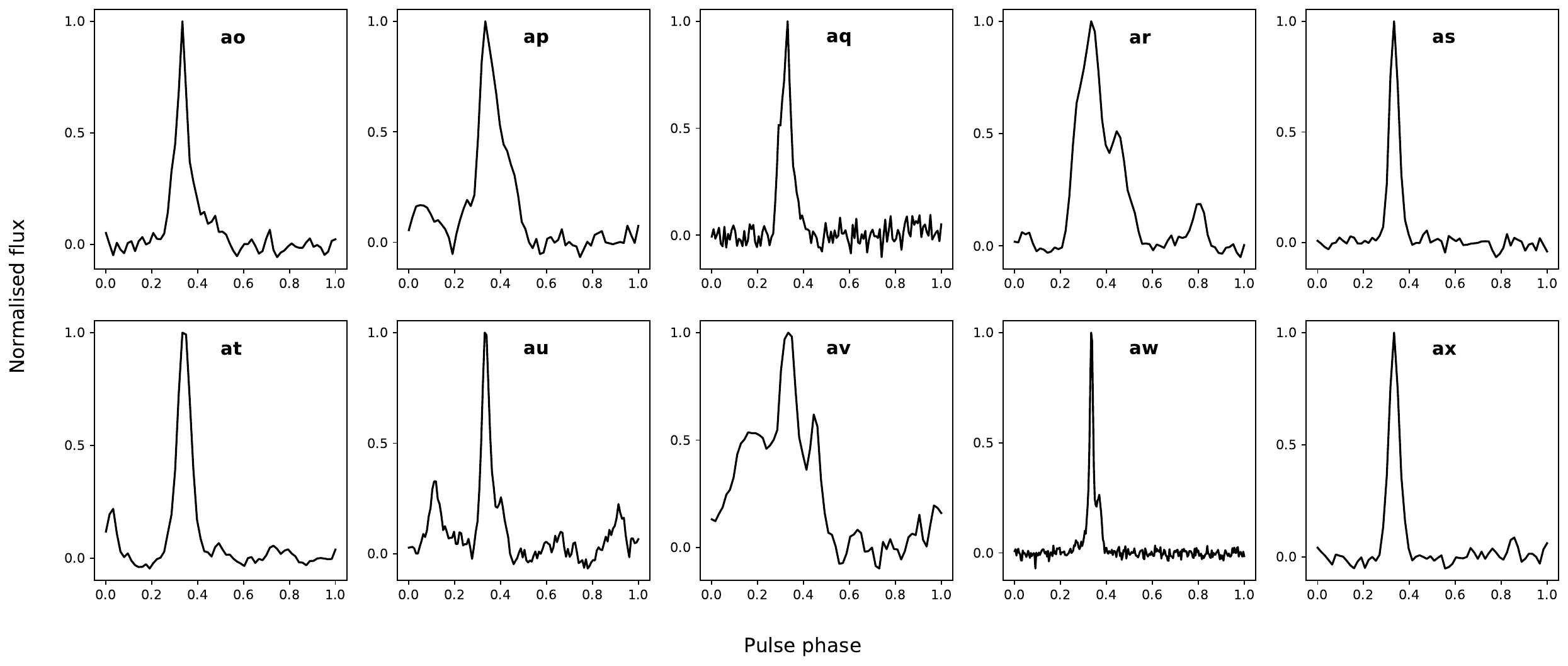}
    \caption{Integrated pulse profiles of all the newly discovered pulsars in Ter5. These plots are obtained after summing together individual profiles from different epochs and aligning the profiles with respect to a reference template profile.}
    \label{fig:Ter5_profiles}
\end{figure*}

We have so far completed the searches and candidate viewing for all beams that lie within the core radius of Ter5 as well as beams that were placed on the positions of known pulsars. 
This totals to roughly 45 beams each for Epoch 1 and 2 (Table \ref{tab:list_observations}). Based on the search strategy described in Section \ref{subsec:search_strategy}, we have confirmed ten new pulsars in Ter5. Nine of these discoveries are in binary systems. Figure \ref{fig:Ter5_profiles} shows a collage of pulse profiles for each of these discoveries. We now give a detailed description of each of the ten Ter5 discoveries.

\subsection{Ter5ao}
\label{subsec:Ter5ao}

Ter5ao (PSR J1748$-$2446ao) was the first Ter5 pulsar to be discovered using data from Epoch 1 of  the TRAPUM GC pulsar survey. It was found with no hint of acceleration. On conducting a refined search at a single DM trial and lowering the threshold for the Fourier significance by a factor of 2, the pulsar was redetected in Epoch 2. The change in barycentric period confirmed the binary nature of the system. Based on these two detections, the GBT data were searched for more detections  after refining the DM obtained from the MeerKAT detections.

The majority of the detections showed a negative acceleration similar to Epoch 1 indicating that the pulsar possibly spends more time behind the companion than in front of it. This suggested that the orbit could be significantly eccentric. Using multiple detections from the GBT data, an orbital solution was derived  using \texttt{fitorb.py}. The initial estimated orbital parameters were $\Pb \simeq 57.55$ d, $x \simeq 62.3$\,lt-s\ and $e = 0.32$. Based on these orbital parameters, the minimum companion mass was estimated (using eq.~\ref{eqn:kepler_third_law}) to be $M_{\rm c, min} \simeq 0.7\, \msun$.
Using this ephemeris, we were able to obtain many more detections over 19 years of GBT data and generated more than 300 topocentric TOAs (including MeerKAT and GBT data). We manually phase connected these TOAs and the final timing solution is provided in Table \ref{tab:eccentric}.

The long timing baseline of 19 years enabled significant measurements of several important astrophysical parameters. Firstly, we were able to measure proper motion in right ascension ($\mu_{\alpha} = -1.6(2)$ \pmunit). Given that Ter5 is located close to the ecliptic ($b$ = 1.68 deg.), constraining the proper motion in declination via timing is difficult. We were also able to get significant detections of spin frequency derivatives up to the fourth order (see Table \ref{tab:eccentric}). Using our measurement of  $\omegadot = 0.0000562 \pm 0.000002\, \rm deg\, yr^{-1}$ and Equation \ref{eqn:mtot_from_omegadot}, we derive  $\Mtot = 3.154(17)\, \msun$. Combining this with the mass function in Equation \ref{eqn:kepler_third_law}, we can additionally constrain $\Mp < 2.23\, \msun$ and $\Mc > 0.92\, \msun$.

If the system has a low orbital inclination, the masses of the pulsar and the companion would be closer to each other and the system would be a double neutron star (DNS) system. However, if the system is nearly edge-on, the pulsar mass could be high ($> 2\, \msun$). This would be evident if a Shapiro delay signature were to be detected in the data. We thus carried out an observing campaign with 5 epochs between 26 Jun 2023 and 07 July 2023 since the time of superior conjunction was expected on 30 Jun 2023. We did not detect a Shapiro delay, so no precise estimates of the individual masses can be made at the moment. We defer the discussion for the Shapiro delay constraints and individual mass measurements to section~\ref{sec:masses}. 

\begin{figure*}
    \centering
    \includegraphics[width=\textwidth]{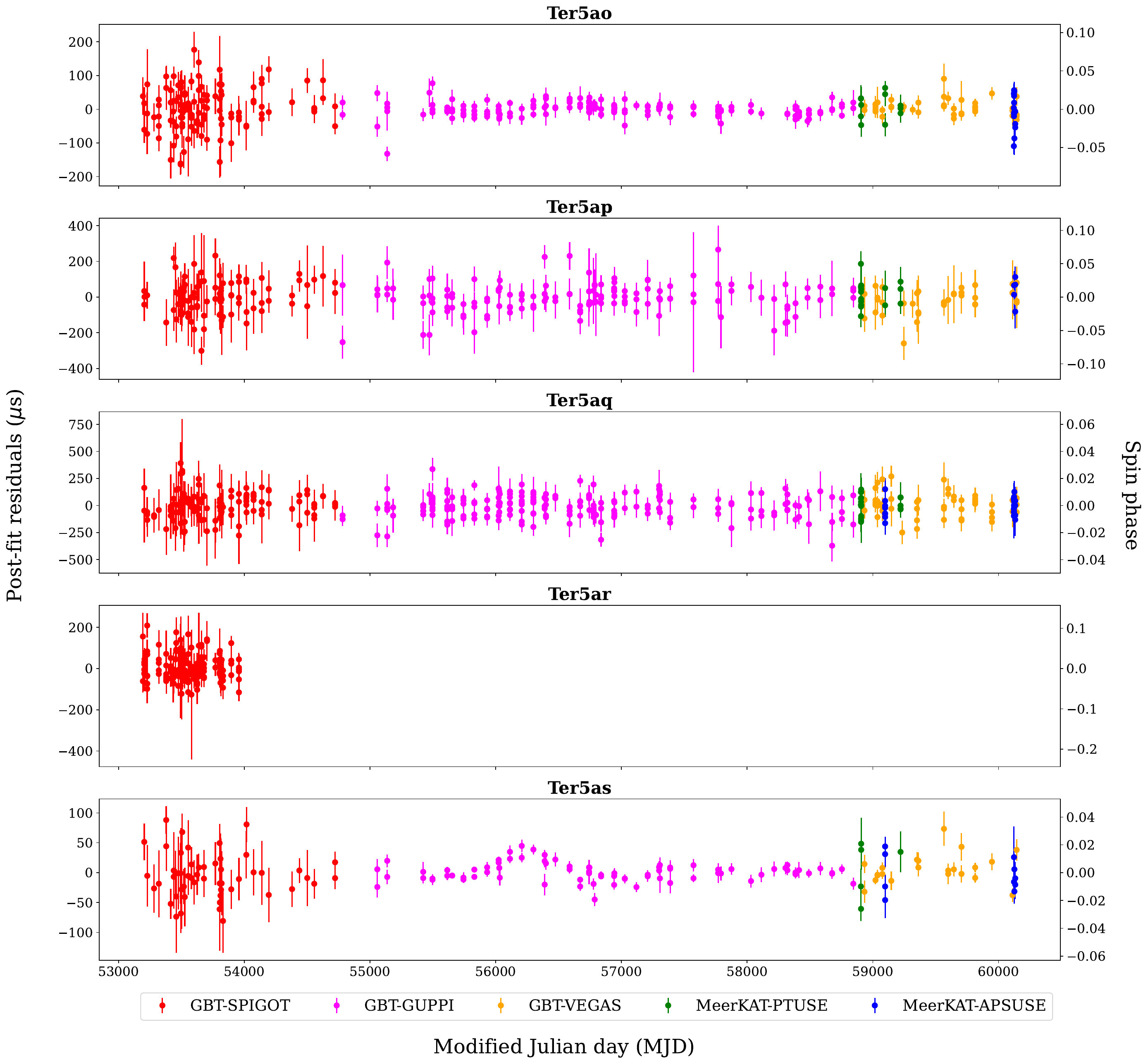}
    \caption{Post-fit residuals after fitting a timing model to the times of arrival (TOAs) obtained from GBT and MeerKAT data spanning about 19 years. Each colour represents TOAs obtained from different back-ends of both the telescopes (as shown in the legend). The secondary y-axis also shows the corresponding residuals as a fraction of the rotational spin phase. These residuals are shown for Ter5ao, Ter5ap, Ter5aq, and Ter5as respectively. The residuals for Ter5ar is depicted here for only 2 years with phase connection.}
    \label{fig:Ter5_residuals_1}
\end{figure*}

\begin{figure*}
    \centering
    \includegraphics[width=\textwidth]{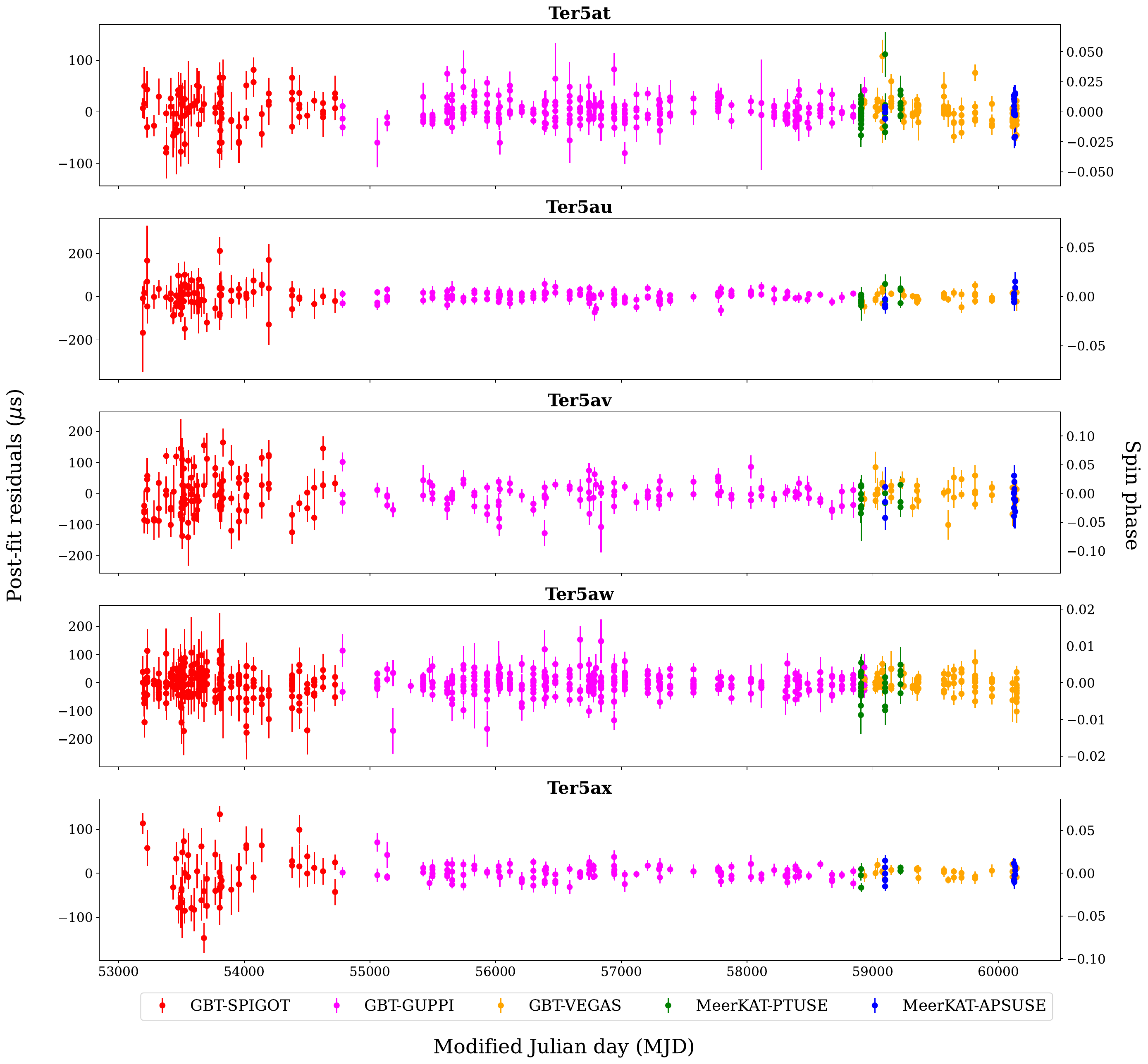}
    \caption{Post-fit residuals as shown in Figure \ref{fig:Ter5_residuals_2} but for the remaining four pulsars with long-term timing solutions. These are Ter5at, Ter5au, Ter5av, Ter5aw and Ter5ax.}
    \label{fig:Ter5_residuals_2}
\end{figure*}

We also obtained a 2.7-$\sigma$ detection of $\xdot = (4.4 \pm 1.6) \times 10^{-14}$ implying
a maximum inclination of $50(9)\, \deg$ (based on Equation \ref{eqn:i_constraint_from_xdot}). If this is confirmed more precisely with continued timing, it will rule out the largest pulsar masses.
We were also able to measure the orbital period derivative $\Pbdot = (164.7 \pm 7.1) \times 10^{-12}\, \rm s\, s^{-1}$ which is consistent with the predicted value from the cluster acceleration alone ($166.38\, \times\, 10^{-12}\, \rm s\, s^{-1}$).
Using Equation~\ref{eqn:p1_pbdot_relation}, we determined the intrinsic spin period derivative to be $\Pdot = 1.13 \times 10^{-20} \rm s\, s^{-1}$  which is consistent with other MSPs. The post-fit residuals for Ter5ao and other pulsars discussed below are depicted in Figures \ref{fig:Ter5_residuals_1} and \ref{fig:Ter5_residuals_2}   

\begin{table*}
\caption[]{Timing parameters for the pulsars Ter5ao, Ter5ap, Ter5au, and Ter5ax as obtained from fitting the observed TOAs with TEMPO. These pulsars are in eccentric orbits. The companion mass is calculated assuming a pulsar mass of $1.35\, \msun$. Numbers in parentheses represent 1-$\sigma$ uncertainties in the last digit. The proper motion in declination value was fixed to $\mu_{\delta}$ = -5.243  mas yr$^{-1}$ based on \cite{Baumgardt_Vasiliev_2021} in cases where it was not measured via timing.}
\begin{center}{\scriptsize
\setlength{\tabcolsep}{6pt}
\renewcommand{\arraystretch}{1.4}
\begin{tabular}{l l l l l}
\hline
Pulsar                                                                         &   J1748$-$2446ao &    J1748$-$2446ap                                                              &  J1748$-$2446au                                                              &   J1748$-$2446ax                                                            \\
\hline\hline
Right Ascension, $\alpha$ (J2000)                                     \dotfill &   17:48:05.14642(7)  &  17:48:05.1291(2)                                                     & 17:48:04.90740(9)                                                            & 17:48:04.46628(6)                                                                          \\
Declination, $\delta$ (J2000)                                         \dotfill &   $-$24:46:36.14(3)  &  $-$24:46:29.8(1)                                                       & $-$24:46:47.75(4)                                                             &  $-$24:46:52.79(2)                                                                            \\
Proper Motion in $\alpha$, $\mu_\alpha$ (mas yr$^{-1}$)               \dotfill &   $-$1.6(2)          &  $-$2.8(5)                                                 &   $-$2.2(3)                                                              &   $-$2.7(2)                                                                     \\
Proper Motion in $\delta$, $\mu_\delta$ (mas yr$^{-1}$)               \dotfill &    $-$5.243          &  $-$5.243                                                   &    $-$5.243                                                              &   $-$18(6)                                             
                       \\
Spin Frequency, $f$ (Hz)                                        \dotfill &   439.68050153408(3)        &    267.04459254652(3)                                         &   219.86646769721(1)                                                     &   514.53606048707(2)                                                       \\
1st Spin Frequency derivative, $\dot{f}$ (Hz s$^{-1}$)                \dotfill &   $-$1.68147(2)$\times 10^{-14}$ & $-$2.189307(9)$\times 10^{-14}$                    &   5.1627(1)$\times 10^{-15}$                                            &   2.52821(2)$\times 10^{-15}$                                               \\
2nd Spin Frequency derivative, $\ddot{f}$ (Hz s$^{-2}$)               \dotfill &   2.1(7)$\times 10^{-26}$        & 1.57(2)$\times 10^{-25}$                                       &   1.93(9)$\times 10^{-25}$                                              &   $-$7.81(3)$\times 10^{-25}$                                             \\
3rd Spin Frequency derivative, $\dddot{f}$ (Hz s$^{-3}$)                \dotfill &   7(3)$\times 10^{-35}$        & --                                &   -8(2)$\times 10^{-35}$                                                 &   10(3)$\times 10^{-35}$                                              
\\
4th Spin Frequency derivative, $\ddddot{f}$ (Hz s$^{-4}$)               \dotfill &   4(1)$\times 10^{-42}$        &  --                                      &   --                                                                     &   3.4(8)$\times 10^{-42}$                                            
      \\
Reference Epoch (MJD)                                                 \dotfill &   56943.753547               &    56674.129312                                        &   56668.690131                                                           &   56668.717373                                                           \\
Start of Timing Data (MJD)                                            \dotfill &   53193.239                  &    53204.063                                        &   53193.201                                                              &   53193.239                                                              \\
End of Timing Data (MJD)                                              \dotfill &   60144.204                  &    60144.196                                        &   60144.179                                                              &   60144.196                                                              \\
Dispersion Measure, DM (pc cm$^{-3}$)                                 \dotfill &   238.205(4)                 &    239.74(6)                                        &   238.08(1)                                                             &   235.449(2)                                                             \\
Solar System Ephemeris                                                \dotfill &   DE440                      &    DE440                                        &   DE440                                                                  &   DE440                                                                  \\
Terrestrial Time Standard                                             \dotfill &   TT(BIPM)                   &    TT(BIPM)                                        &   TT(BIPM)                                                           &   TT(BIPM)                                                               \\
Time Units                                                            \dotfill &   TDB                        &    TDB                                       &   TDB                                                                    &   TDB                                                                    \\
Number of TOAs                                                        \dotfill &   341                        &    298                                      &   279                                                                    &   222                                                                    \\
Residuals RMS ($\mu$s)                                                \dotfill &   20.09                      &    70.33                                        &   22.46                                                                  &   16.70                                                                  \\
$S_{1284}$ (mJy)                                                      \dotfill &   0.012                      &     0.015                                        &  0.012                                                                    &   0.008                                                                   \\
$L_{1284}$ (mJy kpc$^2$)                                              \dotfill  &  0.524                      &     0.647                                         & 0.548                                                                     & 0.376                                                                     \\
Angular offset from nominal cluster centre (arcmin)                                   \dotfill  &  0.156 & 0.253      & 0.053 &  0.161\\ 

\hline
\multicolumn{4}{c}{Binary Parameters}  \\
\hline\hline
Binary Model                                                          \dotfill &   DDFWHE                   &   DDFWHE                                                &   DDFWHE                                                                     &   DD                                                                    \\
Projected Semi-major Axis, $x_{\rm p}$ (lt-s)                         \dotfill &    62.313928(6)            &   13.20131(3)                                             &  6.545721(6)                                              &   14.329676(1)                                                            \\
Orbital Eccentricity, $e$                                             \dotfill &   0.32488898(8)            &   0.905186(4)                                           &   0.025695(1)                                                                      &   9.1542(1)$\times 10^{-3}$                                                                      \\
Longitude of Periastron, $\omega$ (deg)                               \dotfill &   245.29053(1)             &   285.14(4)                                           &   265.935(1)                                                                      &   187.072(1)                                                                      \\
Epoch of passage at Periastron, $T_0$ (MJD)                           \dotfill &   57384.379457(2)          &   56674.842648(3)                                           &   56668.18017(9)                                                        &  56654.2979(1)                                                       \\
Orbital Period, $P_b$ (days)                                          \dotfill &   57.55567566(2)           &   21.38817354(3)                                           &   5.9794622(1)                                                        &   30.208838(1)                                                                     \\
Rate of periastron advance, $\dot{\omega}$ (deg/yr)                   \dotfill &   5.60(3)$\times 10^{-4}$  &   1.058(2)$\times 10^{-2}$                                           &   0.0151(3)                                                                     &   2.4(3)$\times 10^{-3}$                                                                     \\
Orbital Period derivative, $\dot{P}_{\rm b}$ (10$^{-12}$ s s$^{-1}$)  \dotfill &   165(9)                   &    130(21)                                          &   $-$16(1)                                                              &   -13(9)                                                                       \\
Einstein Delay, $\gamma$ (s)                                          \dotfill &   0.00728                  &    -0.0048                                          &    --                                                            &       --          \\
Rate of change of projected semi-major axis  $\dot{x}$                \dotfill  &   4(2)$\times 10^{-14}$   &  -- & -- & -- \\ 
Orthometric amplitude of Shapiro delay, $h_3$ (\textmu s)             \dotfill &   2(1) $\times 10^{-6}$    &    --                                          &   --                                                             &  --   \\ 
Relativistic deformation of the orbit, $\delta_{\theta}$ (10$^{-6}$)  \dotfill &   0.231144                 &      0.346681                                              &   --                                                              &   --                                                                     \\
Relativistic deformation of the orbit, $\delta_{\rm r}$ (10$^{-6}$)   \dotfill &   0.2130047                &     0.3091813                                                &   --                                                              &  --                                                                     \\
Total Mass, $\Mtot$ ($\msun$)                                           \dotfill &   3.17(4)                &      1.997(6)                                           &   1.8(1)                                                                    &    7(1)                                                                    \\
\hline
\multicolumn{4}{c}{Derived Parameters}  \\
\hline\hline
Spin Period, $P$ (s)                                                  \dotfill &   2.2743787739299(1)$\times 10^{-3}$ &  3.7446929385989(5)$\times 10^{-3}$                                   &   4.5482151529223(2)$\times 10^{-3}$                                    &   1.9434983772464(3)$\times 10^{-3}$                                      \\
1st Spin Period derivative, $\dot{P}$ (s s$^{-1}$)                    \dotfill &   8.6979(1)$\times 10^{-20}$         &  3.07000(1)$\times 10^{-19}$                                &   $-$1.06797(2)$\times 10^{-19}$                                         &   $-$9.5495(7)$\times 10^{-21}$                                           \\
Mass Function, $f(M_{\rm p})$ ($\msun$)                       \dotfill &   7.8414183(8) $\times 10^{-2}$              &  5.403(3) $\times 10^{-3}$                                   &   8.422300(7)$\times 10^{-3}$                                    &   3.461973(1)$\times 10^{-3}$                                                   \\
Minimum companion mass, $M_{\rm c, min}$ ($\msun$)            \dotfill &   0.688                                      &     0.238                         &   0.282                                                   &   0.203                                                                    \\
Median companion mass, $M_{\rm c, med}$ ($\msun$)             \dotfill &   0.831                                      &     0.281                         &   0.332                                                  &   0.237                                                                    \\
Surface Magnetic Field, $B_0$, (10$^{8}$ G)                           \dotfill &   1.6231                             &    4.0532                                   &   4.1628                                                                    &   --                                                                     \\
Intrinsic Spin-down, $\dot{P}_{\rm int}$ (10$^{-20}$ s s$^{-1}$)      \dotfill &   1.1312                             &    4.284                                    &  3.7209                                                              &   --                                                                     \\
Characteristic Age, $\tau_{\rm c}$ (Gyr)                              \dotfill &   3.1854                             &    1.384                                     &   1.936                                                                   &   --                                                                     \\
\hline
\end{tabular} }
\end{center}
\label{tab:eccentric}
\end{table*}

\begin{table*}
\caption[]{Timing parameters for the pulsars Ter5aq, Ter5ar and Ter5at as obtained from fitting the observed ToAs with TEMPO. These pulsars are grouped together given their spider nature. Same assumptions have been made as stated earlier in Table \ref{tab:eccentric}}
\begin{center}{\scriptsize
\setlength{\tabcolsep}{6pt}
\renewcommand{\arraystretch}{1.3}
\begin{tabular}{l l l l}
\hline
Pulsar                                                                         &   J1748$-$2446aq                                                              &  J1748$-$2446ar                                                              &   J1748$-$2446at                                                            \\
\hline\hline
Right Ascension, $\alpha$ (J2000)                                     \dotfill &   17:48:04.7344(2)                                                         & 17:48:04.6141(5)                                                            & 17:48:05.36261(3)                                                                          \\
Declination, $\delta$ (J2000)                                         \dotfill &   $-$24:46:42.8(1)                                                           & $-$24:46:45.4(2)                                                             &  $-$24:46:47.07(1)                                                                            \\
Proper Motion in $\alpha$, $\mu_\alpha$ (mas yr$^{-1}$)               \dotfill &   $-$1.7(5)                                                              &   --                                                              &   $-$1.93(9)                                                                     \\
Proper Motion in $\delta$, $\mu_\delta$ (mas yr$^{-1}$)               \dotfill &   $-$5.243                                                              &   --                                                              &   $-$7(3)                                                                     \\
Spin Frequency, $f$ (Hz)                                        \dotfill &   79.859813034659(7)                                                     &   512.082391773(1)                                                     &   456.999686853575(9)                                                       \\
1st Spin Frequency derivative, $\dot{f}$ (Hz s$^{-1}$)                \dotfill &   4.56767(9)$\times 10^{-15}$                                         &   6.94(2)$\times 10^{-14}$                                            &   1.232138(8)$\times 10^{-14}$                                               \\
2nd Spin Frequency derivative, $\ddot{f}$ (Hz s$^{-2}$)               \dotfill &   2.303(5)$\times 10^{-25}$                                                &   $-$3.9(3)$\times 10^{-22}$                                              &  1.583(6)$\times 10^{-25}$                                              \\
3rd Spin Frequency derivative, $\dddot{f}$ (Hz s$^{-3}$)                \dotfill &   $-$6(1)$\times 10^{-35}$                                         &   $-$2.9(4)$\times 10^{-29}$                                                 &   $-$5(1)$\times 10^{-35}$                                              
\\
4th Spin Frequency derivative, $\ddddot{f}$ (Hz s$^{-4}$)               \dotfill &   --                                                &   7(6)$\times 10^{-36}$                                                                    &  --                                            
      \\
Reference Epoch (MJD)                                                 \dotfill &   56674.129312                                                           &   53500.000000                                                           &   56500.000000                                                           \\
Start of Timing Data (MJD)                                            \dotfill &   53204.063                                                              &   53193.201                                                              &   53193.239                                                              \\
End of Timing Data (MJD)                                              \dotfill &   60144.196                                                              &   53957.153                                                              &   60144.199                                                              \\
Dispersion Measure, DM (pc cm$^{-3}$)                                 \dotfill &   238.941(7)                                                             &   238.664(7)                                                             &   239.469(2)                                                             \\
Solar System Ephemeris                                                \dotfill &   DE440                                                                  &   DE440                                                                  &   DE440                                                                  \\
Terrestrial Time Standard                                             \dotfill &   TT(BIPM)                                                               &   TT(BIPM)                                                           &   TT(BIPM)                                                               \\
Time Units                                                            \dotfill &   TDB                                                                    &   TDB                                                                    &   TDB                                                                    \\
Number of TOAs                                                        \dotfill &   422                                                                    &   189                                                                    &   519                                                                    \\
Residuals RMS ($\mu$s)                                                \dotfill &   88.56                                                                  &   51.19                                                                  &   17.70                                                                  \\
$S_{1284}$ (mJy)                                                      \dotfill &   0.017                                                                   &  0.044                                                                    &  0.019                                                                   \\
$L_{1284}$ (mJy kpc$^2$)                                              \dotfill  &  0.730                                                                    & 1.940                                                                     & 0.821                                                                     \\
Angular offset from nominal cluster centre (arcmin)                                   \dotfill  &  0.038     & 0.055 & 0.123\\ 
\hline
\multicolumn{4}{c}{Binary Parameters}  \\
\hline\hline
Binary Model                                                          \dotfill &   BTX                                                                     &   BTX                                                                     &   BTX                                                                    \\
Projected Semi-major Axis, $x_{\rm p}$ (lt-s)                         \dotfill &   0.025864(6)                                                           &  1.498554(8)                                                           &    0.100652(1)                                                           \\
Orbital Eccentricity, $e$                                             \dotfill &   0.0                                                          &        0.0                                                                      &   0.0                                                                      \\
Longitude of Periastron, $\omega$ (deg)                               \dotfill &   0.0                                                           &    0.0                                                                      &   0.0                                                                      \\
Epoch of passage at Periastron, $T_0$ (MJD)                           \dotfill &   59220.484743(6)                                                        &   53495.2744189(5)                                                        &   59097.7002667(4)                                                       \\
Orbital Period, $P_b$ (days)                                          \dotfill &   0.1186466908(2)                                                         &   0.513338728(4)                                                        &   0.2188829327(1)                                                                     \\
Orbital Frequency, $f_{\rm b}$ (s$^{-1}$)                             \dotfill &   9.75507533(2)$\times 10^{-5}$                                          &   2.25466606(2)$\times 10^{-5}$                                          &   5.287791939(2)                                       
 \\
1st Orbital Freq. derivative, $f^{(1)}_{\rm b}$ (s$^{-2}$)            \dotfill &   --                                                                     &  5.1(4)$\times 10^{-19}$                                                  &   5.4(3)$\times 10^{-21}$                                                \\
2nd Orbital Freq. derivative, $f^{(2)}_{\rm b}$ (s$^{-3}$)            \dotfill &   --                                                                     &   9.8(7)$\times 10^{-26}$                                                 &   7(2)$\times 10^{-30}$                                                \\
3rd Orbital Freq. derivative, $f^{(3)}_{\rm b}$ (s$^{-4}$)            \dotfill &   --                                                                     &   $-$1.3(2)$\times 10^{-32}$                                              &   --                                              \\
4th Orbital Freq. derivative, $f^{(4)}_{\rm b}$ (s$^{-5}$)            \dotfill &   --                                                                     &   $-$1.3(2)$\times 10^{-39}$                                              &   --                                             \\
5th Orbital Freq. derivative, $f^{(5)}_{\rm b}$ (s$^{-6}$)            \dotfill &   --                                                                     &   4.2(6)$\times 10^{-46}$                                                 &   --                                                \\
6th Orbital Freq. derivative, $f^{(6)}_{\rm b}$ (s$^{-7}$)            \dotfill &   --                                                                     &   $-$2.8(9)$\times 10^{-53}$                                              &   --                                             \\
\hline
\multicolumn{4}{c}{Derived Parameters}  \\
\hline\hline
Spin Period, $P$ (s)                                                  \dotfill &   1.25219427059248(8)$\times 10^{-2}$                                     &   1.952810750899(4)$\times 10^{-3}$                                    &     2.18818530683239(4)$\times 10^{-3}$                                    \\
1st Spin Period derivative, $\dot{P}$ (s s$^{-1}$)                    \dotfill &   $-$7.16198(6)$\times 10^{-19}$                                           &   $-$2.648(9)$\times 10^{-19}$                                         &   $-$5.89966(4)$\times 10^{-20}$                                           \\
Mass Function, $f(M_{\rm p})$ ($\msun$)                       \dotfill &   1.3197(9)$\times 10^{-6}$                                                   &  1.37117(2)$\times 10^{-2}$                                     &   2.28523(7)$\times 10^{-5}$                                                   \\
Minimum companion mass, $M_{\rm c, min}$ ($\msun$)            \dotfill &   0.013                                                                    &   0.339                                                   &   0.035                                                                    \\
Median companion mass, $M_{\rm c, med}$ ($\msun$)             \dotfill &   0.015                                                                    &   0.401                                                  &   0.041                                                                    \\
\hline
\end{tabular} }
\end{center}
\label{tab:spiders}
\end{table*}

\begin{table*}
\caption[]{Timing parameters for the pulsars Ter5as, Ter5av and Ter5aw as obtained from fitting the observed ToAs with TEMPO. Same assumptions have been made as stated earlier in Table \ref{tab:eccentric}}
\begin{center}{\scriptsize
\setlength{\tabcolsep}{6pt}
\renewcommand{\arraystretch}{1.3}
\begin{tabular}{l l l l}
\hline
Pulsar                                                                         &   J1748$-$2446as                                                              &  J1748$-$2446av                                                              &   J1748$-$2446aw                                                            \\
\hline\hline
Right Ascension, $\alpha$ (J2000)                                     \dotfill &   17:48:04.65947(5)                                                         & 17:48:05.0263(1)                                                            & 17:48:05.41293(6)                                                                          \\
Declination, $\delta$ (J2000)                                         \dotfill &   $-$24:46:40.22(2)                                                           & $-$24:46:43.38(4)                                                             &  $-$24:46:50.04(2)                                                                            \\
Proper Motion in $\alpha$, $\mu_\alpha$ (mas yr$^{-1}$)               \dotfill &   $-$2(1)                                                              &   $-$1.7(2)                                                              &   $-$2.2(1)                                                                     \\
Proper Motion in $\delta$, $\mu_\delta$ (mas yr$^{-1}$)               \dotfill &   $-$12(6)                                                              &   --                                                              &   --                                                                     \\
Spin Frequency, $f$ (Hz)                                        \dotfill &   429.838515818049(9)                                                     &   540.70167993111(2)                                                     &   76.633768847072(2)                                                       \\
1st Spin Frequency derivative, $\dot{f}$ (Hz s$^{-1}$)                \dotfill &   $-$4.72957(1)$\times 10^{-14}$                                         &   1.242661(6)$\times 10^{-14}$                                            &   $-$7.67252(2)$\times 10^{-14}$                                               \\
2nd Spin Frequency derivative, $\ddot{f}$ (Hz s$^{-2}$)               \dotfill &   5.864(7)$\times 10^{-25}$                                                &   $-$6.44(1)$\times 10^{-25}$                                              &   1.44(1)$\times 10^{-26}$                                             \\
3rd Spin Frequency derivative, $\dddot{f}$ (Hz s$^{-3}$)                \dotfill &   7(3)$\times 10^{-35}$                                         &   --                                                 &   -8(3)$\times 10^{-36}$                                              
\\
Reference Epoch (MJD)                                                 \dotfill &   56674.125850                                                           &   56674.105395                                                           &   56500.000000                                                           \\
Start of Timing Data (MJD)                                            \dotfill &   53204.124                                                              &   53204.032                                                              &   53193.239                                                              \\
End of Timing Data (MJD)                                              \dotfill &   60144.128                                                              &   60144.179                                                              &   60144.223                                                              \\
Dispersion Measure, DM (pc cm$^{-3}$)                                 \dotfill &   238.211(1)                                                             &   238.909(2)                                                             &   239.310(2)                                                             \\
Solar System Ephemeris                                                \dotfill &   DE440                                                                  &   DE440                                                                  &   DE440                                                                  \\
Terrestrial Time Standard                                             \dotfill &   TT(BIPM)                                                               &   TT(BIPM)                                                           &   TT(BIPM)                                                               \\
Time Units                                                            \dotfill &   TDB                                                                    &   TDB                                                                    &   TDB                                                                    \\
Number of TOAs                                                        \dotfill &   172                                                                    &   282                                                                    &   706                                                                    \\
Residuals RMS ($\mu$s)                                                \dotfill &   17.26                                                                  &   37.77                                                                  &   27.67                                                                  \\
$S_{1284}$ (mJy)                                                      \dotfill &   0.010                                                                   &  0.015                                                                    &   0.010                                                                   \\
$L_{1284}$ (mJy kpc$^2$)                                              \dotfill  &  0.429                                                                    & 0.679                                                                     & 0.428                                                                     \\
Angular offset from nominal cluster centre (arcmin)                                   \dotfill  &  0.084     & 0.045 &  0.156\\ 
\hline
\multicolumn{4}{c}{Binary Parameters}  \\
\hline\hline
Binary Model                                                          \dotfill &    --                                                                     &   ELL1H                                                                     &   ELL1H                                                                    \\
Projected Semi-major Axis, $x_{\rm p}$ (lt-s)                         \dotfill &    --                                                           &  1.250826(2)                                              &   3.725771(1)                                                            \\
1st Laplace-Lagrange parameter, $\epsilon = e \sin \omega$            \dotfill &    --                                                          &    -0.000099(3)                  &            -0.000014(1) \\
2nd Laplace-Lagrange parameter, $\epsilon = e \cos \omega$            \dotfill &    --                                                          &     -0.000235(4)                   &            -0.000084(1)              \\
Longitude of Periastron, $\omega$ (deg)                               \dotfill &   --                                                           &   203.1(8)                                                                      &   186.9(9)                                                                      \\
Epoch of Ascending Node, $T_\textrm{asc}$ (MJD)                           \dotfill &   --                                                        &   55652.871653(1)                                                        &  59351.17308627                                                        \\
Orbital Period, $P_b$ (days)                                          \dotfill &   --                                                         &   3.381657341(3)                                                        &   0.73135857362(5)                                                                     \\
Rate of periastron advance, $\dot{\omega}$ (deg/yr)                   \dotfill &   --                                                &   --                                                                     &   0.5(1)                                                                     \\
Orbital Period derivative, $\dot{P}_{\rm b}$ (10$^{-12}$ s s$^{-1}$)  \dotfill &   --                                                                 &   $-$10(2)                                                              &   5.92(2)                                                                     \\
Einstein Delay, $\gamma$ (s)                                          \dotfill &   --                                                                &    --                                                            &   0.00000438                                                                     \\
Orthometric amplitude of Shapiro delay, $h_3$ (\textmu s)             \dotfill &  --                                                  &   --                                                                      & 2.1(8) $\times 10^{-6}$
                                                           \\ 
Relativistic deformation of the orbit, $\delta_{\theta}$ (10$^{-6}$)  \dotfill &   --                                                                     &   --                                                              &   3.143218                                                                     \\
Relativistic deformation of the orbit, $\delta_{\rm r}$ (10$^{-6}$)   \dotfill &   --                                                                     &   --                                                              &  2.939424                                                                      \\
Total Mass, $\Mtot$ ($\msun$)                                           \dotfill &   --                                                                 &   --                                                                    &    2.1(6)                                                                    \\
\hline
\multicolumn{4}{c}{Derived Parameters}  \\
\hline\hline
Spin Period, $P$ (s)                                                  \dotfill &   2.32645508301377(5)$\times 10^{-3}$                                   &   1.84944866479313(6)$\times 10^{-3}$                                    &   1.30490776461166(4)$\times 10^{-2}$                                      \\
1st Spin Period derivative, $\dot{P}$ (s s$^{-1}$)                    \dotfill &   2.559829(6)$\times 10^{-19}$                                          &   $-$4.25047(2)$\times 10^{-20}$                                         &   1.306465(3)$\times 10^{-18}$                                           \\
Mass Function, $f(M_{\rm p})$ ($\msun$)                       \dotfill &   --                                                   &   1.83744(1)$\times 10^{-4}$                                    &   0.1038166(4)                                                   \\
Minimum companion mass, $M_{\rm c, min}$ ($\msun$)            \dotfill &   --                                                                    &   0.071                                                   &   0.777                                                                    \\
Median companion mass, $M_{\rm c, med}$ ($\msun$)             \dotfill &   --                                                                    &   0.083                                                  &   0.943                                                                    \\
Surface Magnetic Field, $B_0$, (10$^{8}$ G)                           \dotfill &   --                                                                    &   2.058                                                                    &   10.557                                                                     \\
Intrinsic Spin-down, $\dot{P}_{\rm int}$ (10$^{-20}$ s s$^{-1}$)      \dotfill &   --                                                                     &  2.237                                                              &   8.3411                                                                     \\
Characteristic Age, $\tau_{\rm c}$ (Gyr)                              \dotfill &   --                                                                      &   1.309                                                                   &   2.478                                                                     \\
\hline
\end{tabular} }
\end{center}
\label{tab:other_pulsars}
\end{table*}

\subsection{Ter5ap}
\label{subsec:Ter5ap}

\begin{figure*}
\centering
	\includegraphics[width=0.47\textwidth]{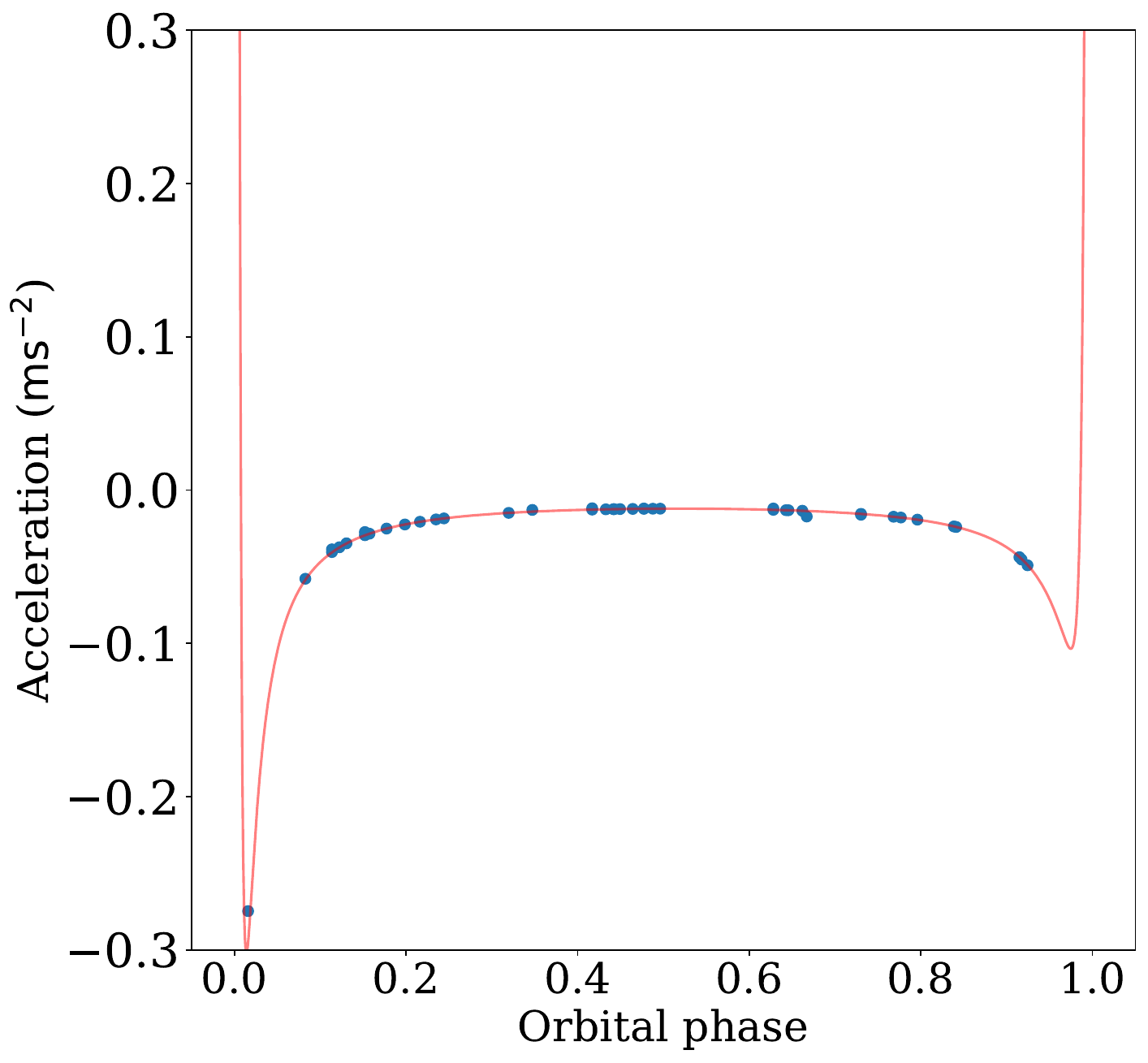}
	\qquad
	\includegraphics[width=0.47\textwidth]{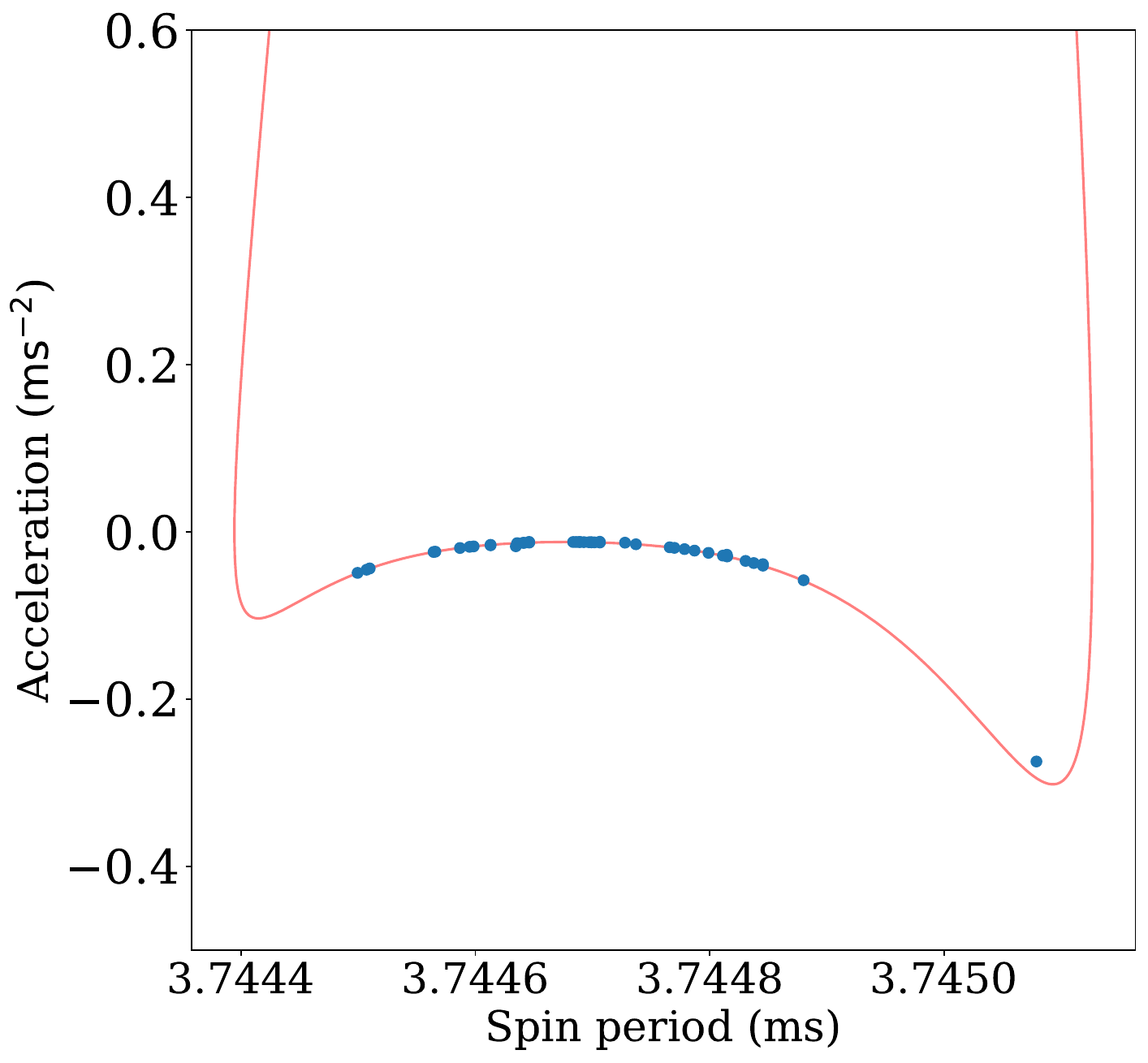}
	\caption{Initial orbital solution for Ter5ap depicted. The left plot shows the line of sight acceleration as a function of orbital phase (which is in turn a function of the mean anomaly). The red line indicates the orbital solution model and the blue dots are data points. Note that the error bars on the data points are smaller than the marking symbol size. All points have a negative acceleration  given that the pulsar spends very little time near periastron (roughly a day out of a 21.3 day orbital period). There are no positive acceleration data points which would better constrain the orbital solution. This is further demonstrated by showing the plot on the right depicting acceleration v/s the observed spin period. One data point clearly does not obey this initial solution.}
  	\label{fig:Ter5ap_orb}
\end{figure*}

Ter5ap (PSR J1748$-$2446ap) was first discovered in Epoch 2. Folding the neighbouring beams revealed multiple detections, suggesting that the candidate was real. Furthermore, it was detected with a significant acceleration of $-$0.280(5)$\rm\, ms^{-2}$. A corresponding refined search of Epoch 1 data revealed no detection. We then attempted to detect the pulsar in GBT data, but only managed to obtain a detection in one epoch. More data were thus needed to obtain an orbital solution. Using data from the campaign between 26 Jun and 07 July 2023, we made 3 more detections. Using multiple iterations of \texttt{fitorb.py} and also aided by the \texttt{binary\_info.py} routine (see Section \ref{subsec:orbital_solution}) we were able to get multiple detections in GBT data that eventually helped in getting an orbital solution with an orbital period of $P_{\rm b} \simeq 21.3 $ d\ and $x \simeq 13.10$\,lt-s\ along with an eccentricity of $e = 0.90$. These parameters imply a highly eccentric orbit where the pulsar has a positive acceleration for roughly 1 day out of the 21.3 day orbit. Figure \ref{fig:Ter5ap_orb} shows the dependence of the observed spin period on the mean anomaly and acceleration as predicted by the orbital solution.

The sparsity of detections from data spanning 19 years proved difficult for long-term phase connection initially despite multiple manual attempts. However, we were able to phase connect the data with a new  technique as mentioned in Section \ref{subsec:timing} (Clark and van Haasteren, in prep). We initially used the GUPPI and VEGAS data to find phase-connected timing solutions using this technique. We found multiple distinct solutions that resulted in similar reduced $\chi^2$ values, all of which extrapolated well to the MeerKAT and SPIGOT TOAs. Upon further inspection, we found that these were related to each other, with spin frequencies separated by $n/P_b$ for small integers $\left|n\right| < 5$ relative to the solution with the least $\chi^2$ value. These were caused by a lack of detections near periastron, around which integer pulsar rotations could be gained or lost without deteriorating the residual $\chi^2$ value. Folding GBT observations closest to periastron with these solutions led to new detections, including in an observation spanning the periastron on MJD 57573, which eventually led to one solution being unambiguously preferred over the others. The final timing solution is presented in Table \ref{tab:eccentric}.

Similar to Ter5ao, we were also able to measure astrometric as well as Post-Keplerian parameters.  We detected proper motion only along the right ascension direction ($\mu_{\alpha} = -2.8(5)$ \pmunit). The timing solution also yielded $\omegadot = 0.010580 \pm 0.000015 \, \rm deg\, yr^{-1}$ implying a total mass of $\Mtot = 1.991 \pm 0.004\, \msun$. Besides this, we also obtained a significant detection of the orbital period derivative $\Pbdot = (130.356 \pm  21.691) \times 10^{-12} \rm \, s\, s^{-1}$ which is consistent with the predicted value ($144.856 \times 10^{-12} \rm \, s\, s^{-1}$). We used Equation~\ref{eqn:p1_pbdot_relation} again to obtain the intrinsic spin period derivative ($\Pdot = 4.284 \times 10^{-20} \rm \, s\, s^{-1}$) and in turn derived the surface magnetic field ($B = 4.05 \times 10^{8}$ G)  and characteristic age ($\tau_c = 1.38$ Gyr) which are all consistent with the MSP population. In Section~\ref{sec:masses}, we discuss the implications of the non detection of  $\gamma$ for Ter5ap on the companion mass.

\subsection{Ter5aq}
\label{subsec:Ter5aq}

Ter5aq (PSR J1748$-$2446aq) was discovered in Epoch 2 and was found in a 1 hour segment at an acceleration of $-2.471(75)\,\rm ms^{-2}$ and a spin period of 12.52\,ms. It was confirmed in Epoch 1 after dedispersing and searching at the DM value from Epoch 2. The Epoch 1 detection showed a different barycentric period and a residual drift in pulse phase indicative of an unaccounted jerk term. These observations suggested a compact binary pulsar. We then tried to obtain detections in other 1 hr segments.  
However, it was detected in just one other segment suggesting that the pulsar could be eclipsing.

Using these two detections we were able to obtain a preliminary orbital solution with $\Pb \simeq 0.12\, \rm d$ and $x \simeq 0.03\, $lt-s, implying $M_{\rm c,min} \sim 0.013 \, \msun$. This solution worked well for Epoch 2 and confirmed that Ter5aq is a black-widow system. The orbital solution yielded multiple detections in data from the corresponding closest GBT observation epochs.

However, this solution resulted in significant phase drift for Epoch 1 suggesting that the orbital period needed improving. Using more detections from GBT, we searched for an improved estimate of $T_0$ using the \texttt{SPIDER\_TWISTER}\footnote{\url{https://github.com/alex88ridolfi/SPIDER_TWISTER}} routine \citep{Ridolfi_2016}. It is useful for detecting pulsars in spider-type systems which experience significant orbital period variability (for well-studied examples, see \citealt{Shaifullah_2016,Ridolfi_2016}), but also for short-period binaries where the orbital period is not known precisely at first.
The routine was used to search for the best $T_0$ value within a limited orbital phase range for the observation epoch of interest. After detecting the pulsar and measuring the local $T_0$, we could further constrain the orbital period by fitting for an integer number of orbits between the different $T_0$ values. This improved the S/N of the detections in Epoch 1.
After generating TOAs for Epoch 1 and Epoch 2 with this ephemeris, we were able to better fit for the orbital parameters. Multiple iterations of this process after including more data points eventually yielded detections from most of the GBT observations across 19 years.

Attempts to manually phase connect the TOAs failed and hence we used the APTB algorithm \citep{Taylor_2024}. Initially, 13 years of data (excluding early GBT data from SPIGOT) were given as an input to APTB. The intricacies involved in these iterations are explained in detail in Section 6 of \citealt{Taylor_2024}. Using this solution, we were able to manually extend the timing baseline to 19 years after inclusion of TOAs from SPIGOT data. The  full timing solution is provided in Table \ref{tab:spiders}.

The final timing solution yielded a significant detection of proper motion only along the right ascension direction ($\mu_{\alpha} = -1.7(5)$ \pmunit) and up to third order spin frequency derivatives (see Table \ref{tab:spiders}). Owing to the circular nature of the system, we were not able to obtain any significant PK parameter measurements despite the long 19-year baseline. In particular, we do not detect the random orbital variability seen in several other black-widows systems. For this reason we can describe the orbital motion using the ELL1 orbital model.

The position of Ter5aq is in close proximity to an X-ray source (CXOU J174804.75-244642.5) with a separation of $\sim$ 0.37 arc-seconds. This source  is reported to be associated with Ter5ab by \cite{Urquhart_2020}. Although Ter5ab is closer to the X-ray source (0.09 arc-seconds), it is isolated. However,  the prominence of X-ray luminosity arising from  eclipsing systems like Ter5aq suggests otherwise and thus warrants further examination.

\subsection{Ter5ar}
\label{subsec:Ter5ar}

Ter5ar (PSR J1748$-$2446ar) was found in a 30 min segment in Epoch 1 at a spin period of 1.95 ms. Inspecting the other segments quickly revealed that the pulsar was found in 7 out of 8 segments in Epoch 1. Using these detections, we were able to  derive a robust orbital solution with $\Pb \simeq 0.51 \, \rm d$\ and $x \simeq 1.5$\,lt-s\ implying a $M_{\rm c,min} \sim 0.34\, \msun$.  Using the derived ephemeris, we were able to get detections with GBT data. We then applied the same strategy as for Ter5aq, by constraining the orbital period using integer orbits between $T_0$ values (provided by \texttt{SPIDER\_TWISTER}) to better constrain the orbital period. Using the new orbital period we were able to obtain a detection in Epoch 2. We then refined the orbital solution further by generating TOAs for both MeerKAT epochs and fitting via \TEMPO. This solution revealed an  eclipse in Epoch 2 confirming that Ter5ar is an eclipsing redback system. We then extended the data baseline to obtain TOAs from all the archival GBT data. 

\begin{figure}
    \centering
    \includegraphics[width=0.48 \textwidth]{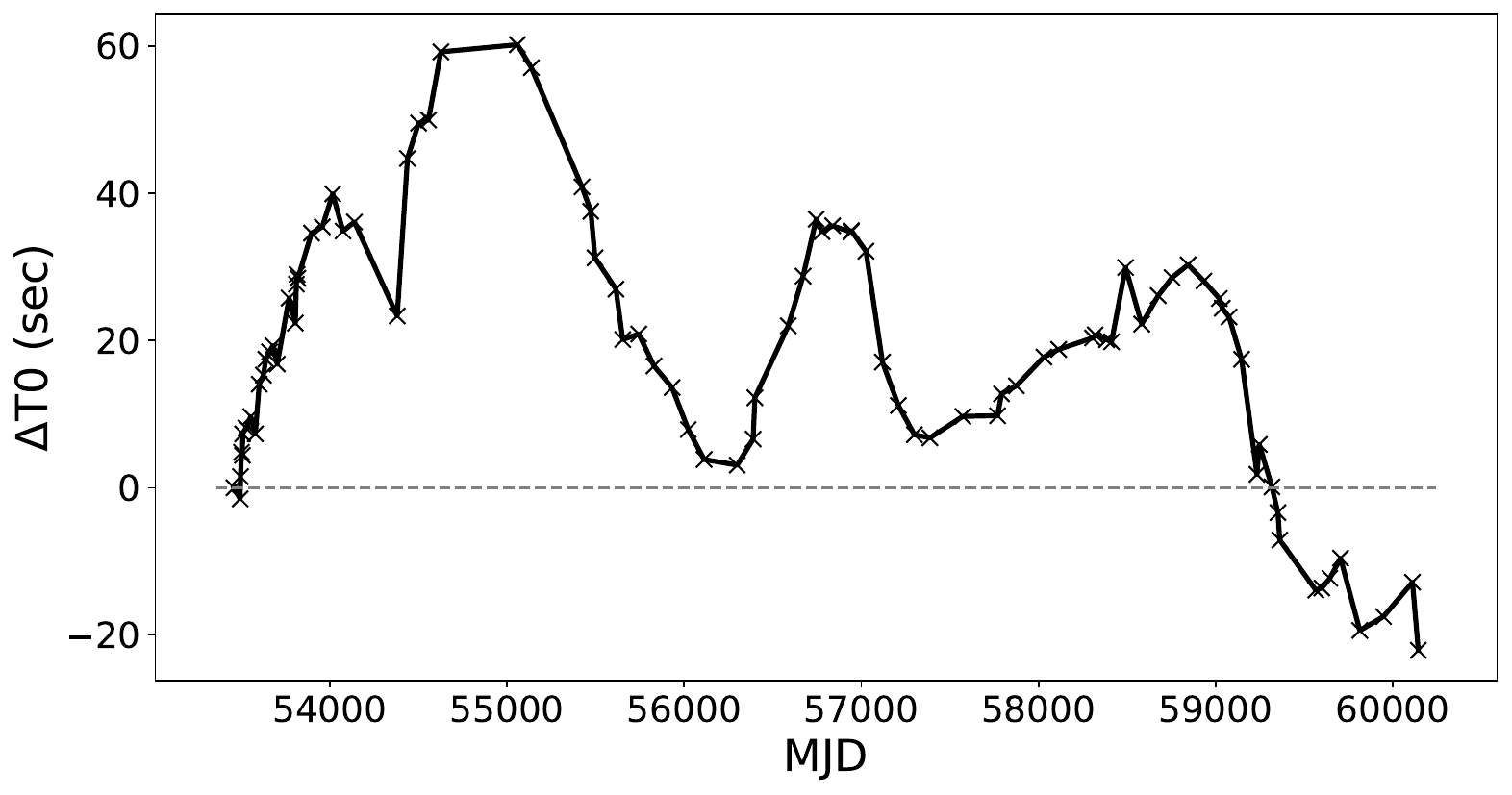}
    \caption{Significant orbital variability in Ter5ar demonstrated. The difference between expected and observed measurement of epoch of periastron (T0) is shown as a function of time in MJD. A stable orbit would show a flat horizontal line but the plot shows the variations ranging from -20 s to 60 s, thus demonstrating the difficulty in timing such systems.}
    \label{fig:Ter5ar_T0_variation}
\end{figure}

When attempting manual phase connection, it was quickly clear that the timing of this system is not trivial. This was not surprising owing to significant long-term changes in the measurement of $T_0$. To demonstrate this, we estimated the $T_0$ value using \texttt{SPIDER\_TWISTER} for every epoch across a 19 year baseline. The variation of $T_0$ across the entire baseline of TOAs obtained is shown in Figure \ref{fig:Ter5ar_T0_variation}. The $\Delta T_0$ that is the difference between the $T_0$ expected using a constant orbital period and the measured $T_0$ can vary by tens of seconds.

A common way to describe such orbits is by using the aforementioned BTX orbital model.
We applied this model to 2 years of GBT data obtained with SPIGOT and were able to manually get a phase connected solution only after applying six orbital frequency derivatives and up to fourth order spin frequency derivatives.  
It was difficult to extrapolate data points beyond 2 years due to the scale of the orbital variability (as shown in Figure \ref{fig:Ter5ar_T0_variation}). Moreover, the cadence of observations after these 2 years also reduced from approximately once a month to once in 3 months. This two year timing solution is given in Table \ref{tab:spiders}. Interestingly, APTB was independently able to phase connect data spanning a year, but the high orbital variability prevented convergence with longer datasets \citep{Taylor_2024}. A deeper analysis of the timing variability across the entire data span and in turn obtaining a long-term timing solution is work in progress.

\citet{Urquhart_2020} conducted deep radio continuum imaging of Ter5 at 2--8 GHz yielding 24 sources where 19 of them could be associated with previously known pulsars and X-ray binary systems. They observed 3 such sources to have a steep spectral index in radio data and a hard X-ray photon index, suggesting a spider-type system with an intra-binary shock. One of these systems, namely VLA-38 was shown to have an X-ray light curve with a periodicity of 12.32 hours similar to Ter5ar.
The position obtained from radio timing ($\alpha = 17\h48\m04\fs6141(5)$; $\delta = -24\degr46\arcmin45\farcs4(2)$) is consistent with the position quoted for VLA-38 in radio imaging and CXOU J174804.63$-$244645.2 from X-ray imaging (as reported in \citet{Urquhart_2020}) all to within 0.6 arc-seconds (see Figure \ref{fig:Ter5_xray}). On fixing the timing position to the VLA-38 source coordinates we were still able to hold phase connection. Considering this along with the matching orbital period from X-ray as well as radio observations, we can unambiguously link Ter5ar with VLA-38.

\subsection{Ter5as}
\label{subsec:Ter5as}

Ter5as (PSR J1748$-$2446as) was discovered in Epoch 1 at a spin period of 2.32 ms. Although it could not be detected in Epoch 2 after searching in a restricted DM range, it was visible in multiple epochs with GBT data. All the barycentric spin periods were consistent with the pulsar being an isolated system. Thus, obtaining more detections across all the epochs was relatively simple compared to the binary pulsars discussed earlier. Consequently, we were able to manually obtain a phase coherent timing solution spanning 19 years.

The timing solution revealed significant proper motion  in right ascension ($\mu_{\alpha} = -2(1)$ \pmunit) as well as in declination ($\mu_{\delta} = -12(6)$ \pmunit). The timing solution also yielded a high first order spin frequency derivative value that is $\dot{f} = -4.7295(1) \times 10^{-14} \,\rm Hz \, s^{-1}$ as well as a significant $\ddot{f}$ term. This could explain a significant drift in spin frequency with time thus reducing the usefulness of stacking across long time span datasets. The full timing solution is given in Table \ref{tab:other_pulsars} and the post-fit timing residuals are plotted in Figure \ref{fig:Ter5_residuals_1}.

\subsection{Ter5at}
\label{subsec:Ter5at}

Ter5at (PSR J1748$-$2446at) was discovered in Epoch 1 in a 30 min segment at a spin period of $2.188\, \rm ms$ and at an acceleration of 0.43(11) $\rm ms^{-2}$. Examining the other 30 min segments revealed three more detections. Two segments showed the pulse to be fading indicating a possible ingress and egress of a radio eclipse and suggesting another spider-type system. Using these detections we were able to derive an initial orbital solution of $P_{\rm b} \simeq 0.22 \, \rm d$\ and $x \simeq 0.1$\,lt-s\ implying a $M_{\rm c,min} \sim 0.03\, \msun$. Folding the entire Epoch 1 time span with this ephemeris revealed a radio eclipse, confirming that Ter5at is an eclipsing black widow system.

We then extracted TOAs from the detections in Epoch 1 and Epoch 2 to get a better constrained timing solution. Using similar strategies to those applied to Ter5aq, we obtained  multiple detections of Ter5at in all the datasets spanning 19 years. We again used APTB  to obtain a fully phase connected timing solution for Ter5at across a 13 year time span (excluding SPIGOT data). We then extracted TOAs for the remaining 6 years to extend the timing baseline to 19 years. Owing to noticeable changes in the residuals from orbital variability, we switched to the BTX timing model and included two orbital frequency derivatives (see Table \ref{tab:spiders}). 

The timing solution yielded a significant value of proper motion ($\mu_{\alpha} = -1.93(9)$ \pmunit; $\mu_{\delta} = -7(3)$ \pmunit) and needed the first three spin frequency derivatives. Assuming that Equations \ref{eqn:pbdot_measured} and \ref{eqn:p1_pbdot_relation} hold, we observe that the expected $\Pbdot$ value from the cluster acceleration is a factor of 3 lower than the measured value suggesting other effects may be at play. Additionally, we observed that Ter5at is located just 0.29 arc-seconds from the X-ray source CXOU J174805.37$-$244646.7 suggesting a possible association, which is supported by the eclipsing nature of the system.     

\subsection{Ter5au}
\label{subsec:Ter5au}

Ter5au (PSR J1748$-$2446au) was discovered in Epoch 1 in the full 4 hour observation at a spin period of 4.54 ms and at an acceleration of $0.113(3) \, \rm ms^{-2}$ and was also independently detected in Epoch 2.

Similar to Ter5ap, Ter5au was initially detected in just a few GBT epochs and required more detections to obtain a reasonable orbital solution. We thus used the 5 follow-up campaign epochs (that is Obs ID 08L-orb to 12L-orb in Table \ref{tab:list_observations}) to search for Ter5au and obtained three detections. Using the MeerKAT detections and the initial GBT detections, we were able to obtain a preliminary orbital solution using \texttt{fitorb.py} yielding $\Pb \simeq 5.97 \, \rm d$ and $x \simeq 6.55 $\,lt-s\ with a hint of eccentricity ($e \sim  0.02$) implying $M_{\rm c,min} \sim 0.28\, \msun$.  Using this orbital solution as an input for \texttt{binary\_info.py}, we were able to get several detections across the GBT data span. Most detections showed the pulses were not drifting in rotational phase. We then generated TOAs for all the data and were able to manually phase connect all the available data. The full timing solution is given in Table \ref{tab:eccentric}.

We were able to detect proper motion in right ascension alone ($\mu_{\alpha} = -2.4(2)\,$\pmunit), and also a significant $\Pbdotobs = (-16.357 \pm 1.186) \times 10^{-12} \, \rm s\, s^{-1}$. Using Equation \ref{eqn:p1_pbdot_relation}, we obtained the intrinsic spin period derivative $\Pdot = 3.72 \times 10^{-20}\, \rm s\, s^{-1}$ which is a typical value of the MSP population.

Furthermore, we obtained a significant detection of $\omegadot = 0.01513 \pm 0.00037 \, \rm deg\, yr^{-1}$ implying a total mass of $\Mtot = 1.83 \pm 0.07\, \msun$. The measured mass function and the nominal value of $\Mtot$ would imply that $\Mp < 1.53 \, \msun$ and $\Mc > 0.30\, \msun$. This indicates that the companion is not a He WD: for the orbital period of this system, \cite{Tauris_Savonije_1999} predict a WD mass of $\sim 0.24 \, \msun$. This means that the companion is more likely to be a relatively light carbon-oxygen (CO) WD. The relatively large mass of the WD companion, the spin of the pulsar and the orbital period make the system look very similar to PSR~J1614$-$2230, which is thought to have evolved via case A Roche lobe overflow (RLO, \citealt{Tauris_Langer_Kramer_2011}) when the donor star (that is the progenitor of the companion) was still in the main sequence stage.

\subsection{Ter5av}
\label{subsec:Ter5av}

Ter5av (PSR J1748$-$2446av) was discovered in Epoch 1 at a spin period of 1.85 ms and an acceleration of $-0.1642(3)\, \rm ms^{-2}$ and was independently detected in Epoch 2. We were then able to get a few detections in GBT data using the DM as a constraint. However, these detections were not enough to get a well constrained orbital solution. We therefore searched the follow-up campaign epochs (similar to the strategy applied for Ter5ap and Ter5au) and obtained three more detections.

Using these GBT and MeerKAT detections, we determined a preliminary orbital solution with \texttt{fitorb.py} of $\Pb \simeq 3.38\, \rm d$\ and $x \simeq 1.25 $\,lt-s\ implying $M_{\rm c,min} \sim 0.07\, \msun$. The \texttt{binary\_info.py} script  was used with the orbital solution to help yield many more GBT detections. Another iteration with \texttt{fitorb.py}, after including the new detections, yielded a low non-zero eccentricity ($e \sim 0.0002$). This pulsar had a broad profile with faint detections across most of the GBT data, resulting in low timing precision and difficulty in obtaining phase connection.

To aid the timing process, we used \texttt{SeeKAT} to get a better position for Ter5av. During one of the five follow-up campaign epochs, we placed seven beams that were hexagonally packed with the central beam centered on the position of the best beam detection in Epoch 1. This beam tiling pattern resulted in a robust position from \texttt{SeeKAT} with sub-arcsecond precision. We generated TOAs for all the data and used the newly obtained position as a starting point for attempting phase connection. This step was integral to successfully phase connect the data manually. The final timing solution is presented in Table \ref{tab:other_pulsars}. 

We were able to detect proper motion in right ascension alone ($\mu_{\alpha} = -$1.7(2) \pmunit). We were also able to detect $\Pbdot = (-10.122 \pm 1.909) \times 10^{-12}\, \rm s\,s^{-1}$ and using Equation \ref{eqn:p1_pbdot_relation}, we obtain the intrinsic spin period derivative value of $\Pdot = 2.23 \times 10^{-20} \rm s\,s^{-1}$ with a derived surface magnetic field of $B = 2.05 \times 10^{8}$ G and characteristic age of $\tau_c = 1.3$ Gyr. We also identified a potential association with X-ray source CXOU J174805.05$-$244643.1 which is within 0.43 arc-seconds of the position of Ter5av. 

\subsection{Ter5aw}
\label{subsec:Ter5aw}

Ter5aw (PSR J1748$-$2446aw) was discovered in Epoch 1 in a 1-hour segment with a spin period of 13.04 ms, an acceleration of 5.77(2) $\rm ms^{-2}$ and confirmed in Epoch 2. Searching 1-hr long data segments across a few epochs of GBT data revealed several detections. These detections helped obtain a starting orbital solution with  $\Pb \simeq 0.73\, \rm d$ and $x \simeq 3.72 $\,lt-s\ implying $M_{\rm c,min} \sim 0.8\, \msun$. With no significant detection of eccentricity, we suspected that the companion is likely a high mass white dwarf. Using this starting orbital solution, we were able to generate TOAs for all GBT data. We then proceeded with manual phase connection and got a fully phase connected timing solution for all the GBT and MeerKAT data. The final timing solution is provided in Table \ref{tab:other_pulsars}.

The 19-year timing solution yielded a significant detection of proper motion ($\mu_{\alpha} = -2.2(1)\,$\pmunit; $\mu_{\delta} = -14(3)\,$\pmunit). Furthermore, we also detected spin frequency derivatives up to the fourth order, similar to Ter5ao (see Table \ref{tab:other_pulsars}). The timing solution also yielded the detection of  $\omegadot = 0.55 \pm 0.11 \, \rm deg\, yr^{-1}$ implying a total mass of $\Mtot = 2.12 \pm 0.61\, \msun$.

We also detected the lowest order orbital period derivative $\Pbdotobs = (5.924 \pm  0.011) \times 10^{-12} \rm \, s\, s^{-1}$ which is slightly lower than the upper limit derived from $\Pdot$ ($6.139 \times 10^{-12} \rm \, s\, s^{-1}$). Using the measured $\Pbdot$ value we obtained estimates for the intrinsic spin period derivative ($\Pdot = 8.341 \times 10^{-20} \rm \, s\, s^{-1}$), surface magnetic field ($B = 1.05 \times 10^{9}$ G) and characteristic age ($\tau_c = 2.48$ Gyr) which are consistent with the population of pulsars which are mildly recycled. 

Finally, we obtained a hint of Shapiro delay in this system: the orthometric amplitude in our DDFWHE solution ($h_3 = 2.13 \pm 0.88 \, \upmu s$) has a low (2.4-$\sigma$) significance.
For this reason, and also because of the relatively low precision of $\dot{\omega}$, no precise masses can be derived yet (see discussion in section~\ref{sec:masses}). However, with its relatively slow spin period and massive WD companion, the system strongly resembles PSR~J1952+2630, a pulsar--massive WD system \citep{Gautam_2022b}.

\subsection{Ter5ax}
\label{subsec:Ter5ax}

Ter5ax (PSR J1748$-$2446ax) was discovered in Epoch 2 in the full 4 hour data span at a spin period of 1.94 ms and an acceleration of 0.005(1) $\rm ms^{-2}$. A search in Epoch 1 revealed no detection. However, a blind search in GBT data revealed a few more detections. But the total number of detections and the wide spacing between the observation epochs was insufficient to solve for a unique orbital solution. We thus searched the data from the follow-up campaign and obtained detections in 4 out of 5 epochs.

Using all the MeerKAT detections and the few GBT detections, we were able to obtain a preliminary orbital solution with  $\Pb \simeq 30.20 \, \rm d$\ and $x \simeq 14.31 $\,lt-s\ implying a $M_{\rm c,min} \sim 0.20\, \msun$, thus suggesting a He WD companion. This preliminary orbital solution served as an input to \texttt{binary\_info.py} and helped obtain many more detections in GBT data. Refitting for the orbital solution yielded a small non-zero eccentricity ($e = 0.0091(2)$). We then used this solution to re-fold all the data back till 2009 and were able manually to phase connect all the data. This solution was then used to generate TOAs from SPIGOT data back till 2004 and we were eventually able to obtain a 19 year phase connected timing solution (  Table \ref{tab:eccentric}).   

We were able to detect significant proper motion along both directions ($\mu_{\alpha} = -$2.7(2) \pmunit; $\mu_{\delta} = -$18(6) \pmunit). We also measured spin frequency derivatives up to the fourth order (see Table \ref{tab:eccentric}). Without including the SPIGOT data, there was a significant detection of $\omegadot = 0.00118 \pm 0.00036\, \rm deg\, yr^{-1}$ implying a very uncertain total mass of $\Mtot = 2.30 \pm \,1.06 \msun$. However, on including the SPIGOT data,  $\omegadot = 0.002411 \pm 0.000266\, \rm deg\, yr^{-1}$ suggesting a total mass of $\Mtot \sim 6.68\, \msun$ if all the contribution to $\omegadot$ is considered to be relativistic. Unless the system has a very low orbital inclination, the contribution to $\omegadot$ most likely comes from additional components apart from GR. Given that the $\omegadot$ detection after including SPIGOT data is at least 5 sigma significant, it points to an unresolved discrepancy.

To further understand this, we write down the relativisitic and classical effects that can cause secular changes in the observed  rate of periastron advance ($\dot{\omega}_{\rm obs}$). This is given by \citep[see e.g.][]{Lorimer_Kramer_2004}:
\begin{equation}
\label{eqn:omega_dot_all_contributions}
\dot{\omega}_{\rm obs} =  \dot{\omega}_{\rm rel} + \dot{\omega}_{\rm PM} + \dot{\omega}_{\rm SO} + \dot{\omega}_{\rm third},
\end{equation}
where $\dot{\omega}_{\rm rel}$ corresponds to the contribution from the relativistic effects within the orbit (eq.~\ref{eqn:mtot_from_omegadot}),  $\dot{\omega}_{\rm PM}$ is the contribution from proper motion,  $\dot{\omega}_{\rm SO}$ from spin-orbit coupling and  $\dot{\omega}_{\rm third}$ is the contribution from a  secondary outer companion. 

However, the pulsar is faint in the less sensitive SPIGOT data and  RFI could be be causing unexpected issues. Additionally, the measurement of $\omegadot$ before and after adding the SPIGOT data are barely compatible at a 2-$\sigma$ level. The timing data for this pulsar is currently being further investigated to understand this better.   

\section{Mass estimates}
\label{sec:masses}

\subsection{Bayesian map}

In order to investigate the mass constraints on the systems where we measure $\omegadot$ (that is Ter5ao, ap, au, and aw), we made a map of the quality of fit (the $\chi^2$) for a grid of $\Mtot$ and $\cos i$. This uniform grid amounts to a uniform prior on the total mass, which is constrained by the detection of omega-dot via Equation \ref{eqn:mtot_from_omegadot}, and an isotropic prior for the orbital axis.
For each point in this grid, we calculate $i$ and then, using Equation ~\ref{eqn:kepler_third_law}, we calculate $\Mc$. 

We used the DDGR timing model, which is similar to the DD model but assumes that GR is the  correct theory of gravity \citep{Damour_Deruelle_1986}. For each point in the $\Mtot$-$\cos i$ grid, the $\Mtot$ and $\Mc$ values are fixed and all other parameters are allowed to vary. The resulting $\chi^2$ describes how well the values of $\Mtot$ and $\Mc$ can describe the timing data.

\begin{figure*}
    \centering
    \includegraphics[width=.6\textwidth]{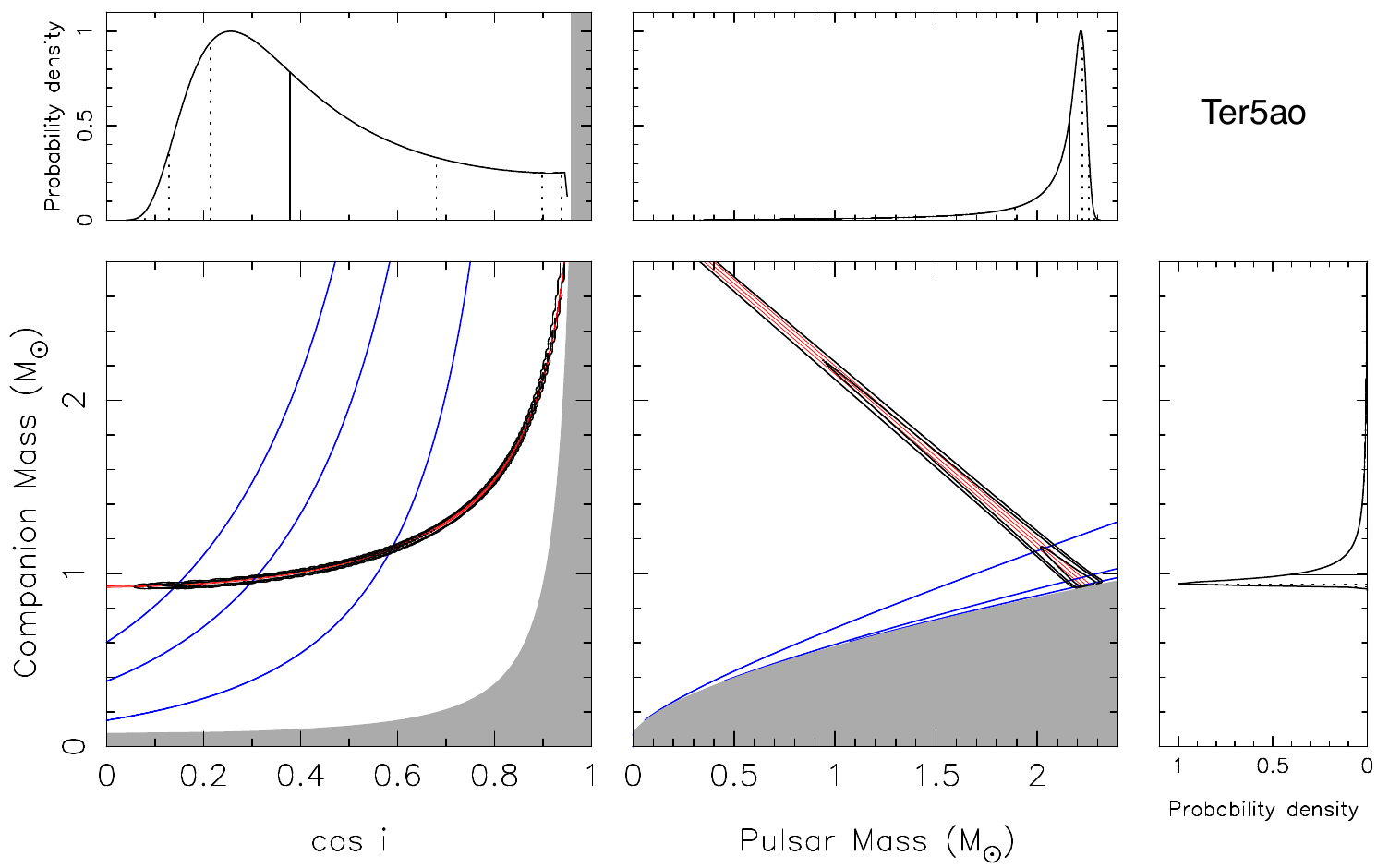}
    \includegraphics[width=.6\textwidth]{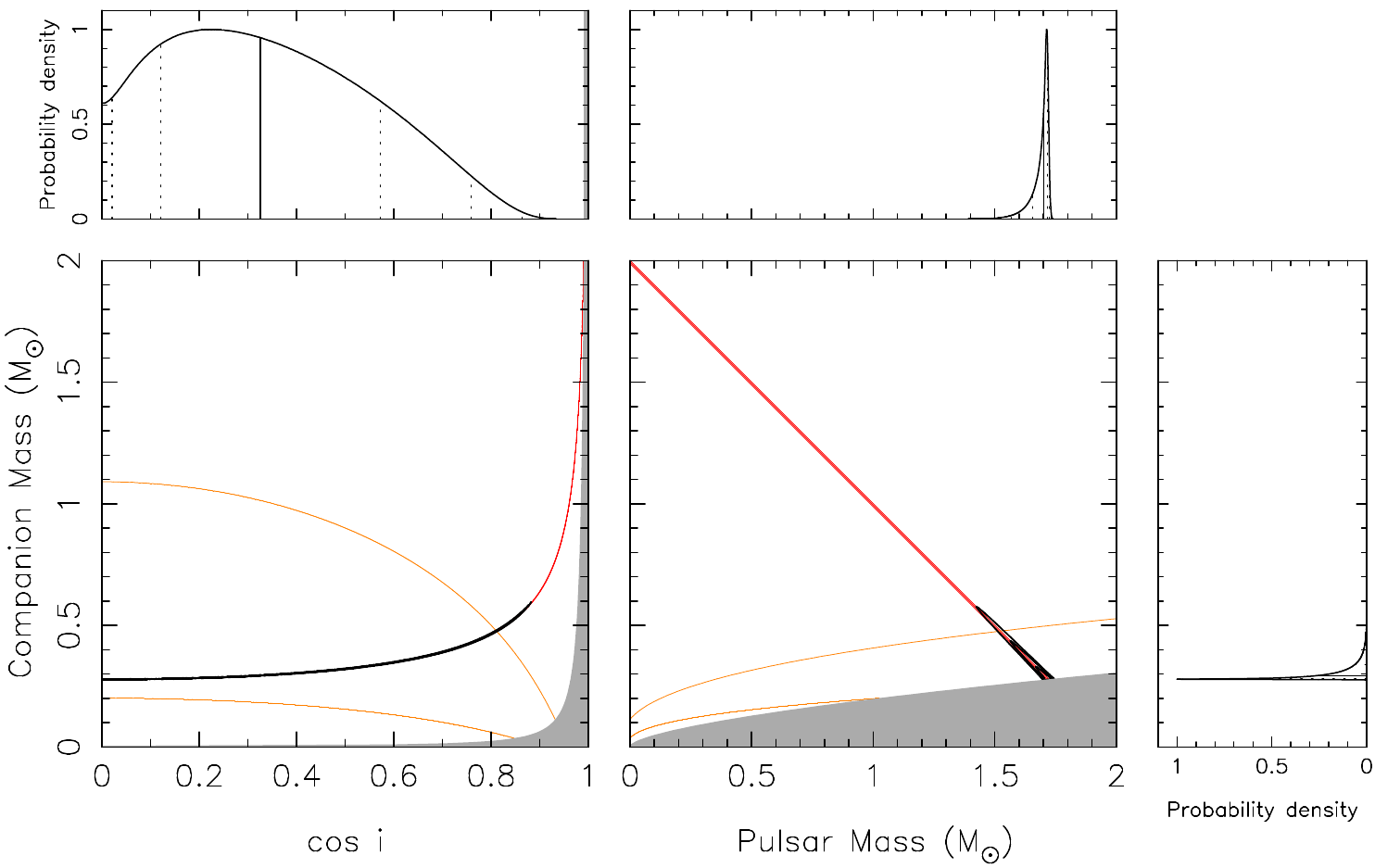}
    \includegraphics[width=.6\textwidth]{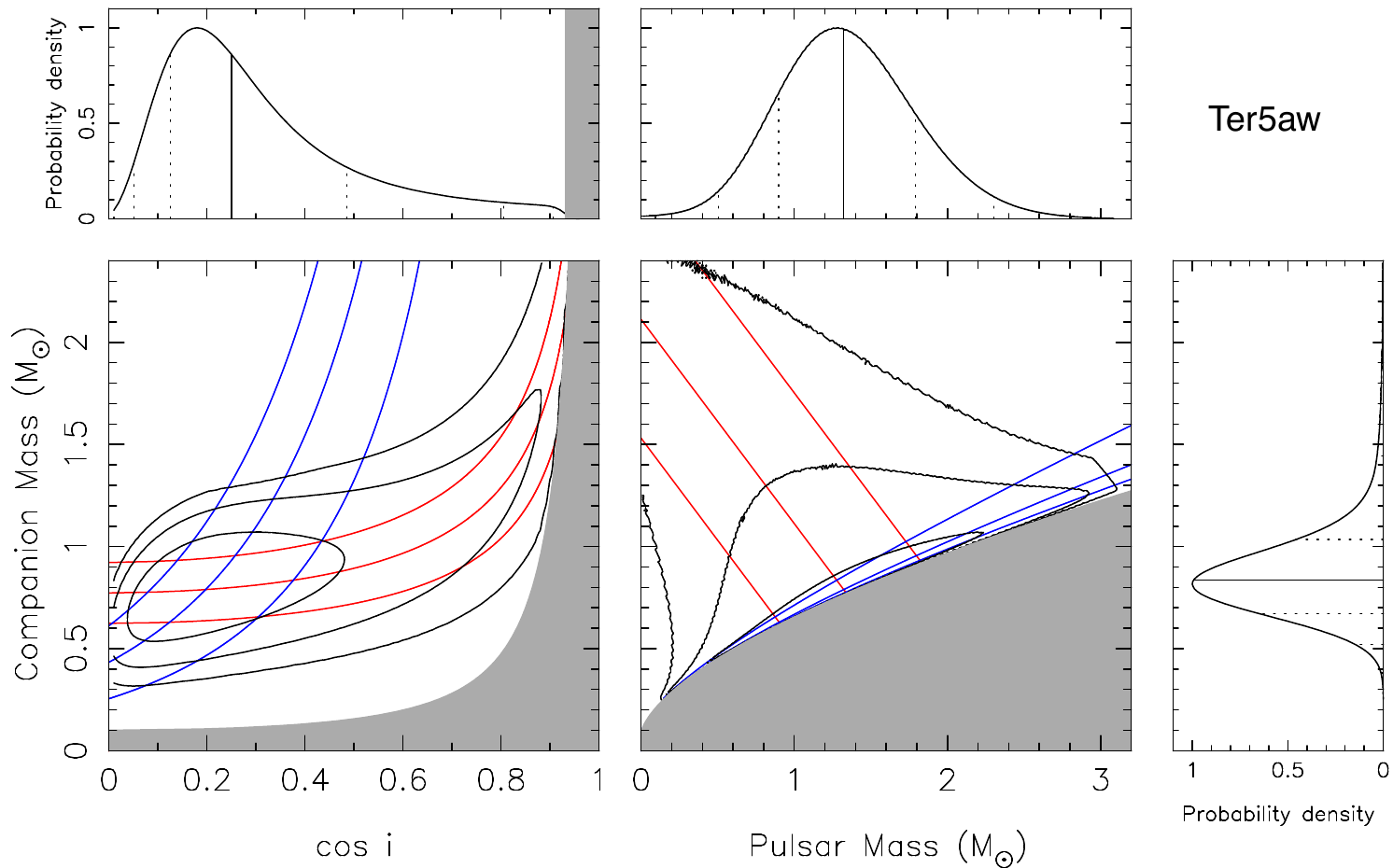}
    \caption{Mass-mass diagrams for Ter5ao (top), Ter5ap (middle), and Ter5aw (bottom). The main panels show $\Mc$ as a function of $\cos i$ (left) and $\Mp$ (right). In the left panels, the regions marked in gray are excluded because they imply a negative pulsar mass. For the panels on the right, the gray regions are constrained by the mass function and  $\sin i < 1$. 
    The black contours include 68.3, 95.4 and 99.7\% of all probability in the 2-D joint posterior probability distribution functions (pdfs) of each panel. 
    The red lines indicate constraints derived from the nominal values of the total mass derived from the nominal value of $\omegadot$ and the $\pm$1-$\sigma$ uncertainties of its measurement. The minimum companion masses for Ter5ao and Ter5ap are constrained by $\dot{\omega}$ and the limit of $\sin i < 1$ giving $\Mc > 0.8 \msun$ and $\Mc > 0.278 \msun$ respectively, this results in $\Mp < 2.23 \msun$ and $\Mp < 1.72 \msun$ respectively. 
    The blue lines represent the constraints derived from the nominal and $\pm 1$-$\sigma$ limits of the orthometric amplitude of the Shapiro delay, $h_3$. The orange lines designate the 2 and 3-$\sigma$ upper limits derived for Ter5ap. The top panels depict the normalised 1-D pdfs for $\cos i$, $\Mp$ and, on the side, $\Mc$. The medians of the pdfs are depicted by the solid black lines, the dotted lines indicate successive $\pm 1$ and $2$-$\sigma$ equivalent percentiles around the median.}
    \label{fig:mass_mass_diagrams}
\end{figure*}

After this stage, we use the Bayesian technique described by \cite{Splaver_2002}. The likelihood is calculated from the $\chi^2$
using
\begin{equation}
 p( X\, |\, \Mtot, \cos i) = \frac{1}{2}e^{-\frac{\chi^2 - \chi^2_{\rm min}}{2}},
\end{equation}
where $X$ represents the  set of all TOAs with uncertainties from all epochs and  $\chi^2_{\rm min}$ is the smallest value of $\chi^2$ in the whole grid (corresponding, therefore, to the best fit, which will therefore have the highest probability density). Using Bayes' theorem, we then derive 2-dimensional joint posterior probability distribution functions (pdfs) for $\Mc$ and $\cos i$ (see contours in left main panels in Fig.~\ref{fig:mass_mass_diagrams}), for $\Mc$ and $\Mp$ (right main panels in Fig.~\ref{fig:mass_mass_diagrams}), and 1-D pdfs for $\Mc$, $\Mp$, $\cos i$, (see lateral panels) and $\Mtot$.

\subsection{Results}

The $\chi^2$ maps confirm the fact mentioned earlier that there is still no significant detection of the Shapiro delay for any of the systems: indeed, in the top left side panels of the distributions obtained for both Ter5ao and Ter5aw , we see significant probability tails for the lower inclinations.
For Ter5aw, all mass values are at this stage too imprecise to be astrophysically useful: the companion mass is consistent with what is expected for a CO WD \citep[see e.g.][]{Shamohammadi_2023}, but the mass is too uncertain for any firm conclusions.

Our $\chi^2$ map for Ter5au (not displayed in Fig.~\ref{fig:mass_mass_diagrams}) confirms that, apart from the well-measured $\Mtot$, there are no additional constraints on cos $i$, which has a nearly flat pdf (the median for $i$ is 60 deg, as we would expect from the assumed prior).  

However, for Ter5ap, there is a significant decrease of the probability for lower inclinations.
This is caused not by a detection of the Shapiro delay, by the small value of the Einstein delay measured for this system, $\gamma = -4.8 \pm 4.4 \, \rm ms$, which excludes large companion masses, as we can see in the middle plot of Fig.~\ref{fig:mass_mass_diagrams}. This small measurement of $\gamma$ cannot be due to a measurement of $\dot{x}_{\mu}$ that is being interpreted as $\gamma$.
Evaluating eq.~\ref{eqn:xdot_contribution} and eq. 25 of \cite{Ridolfi_2019_NGC1851A}, we find that, for Ter5ap; $\dot{x}_{\gamma} >> \dot{x}_{\mu}$, which implies that the effect of $\gamma$ should be dominant.

The opposite is true for Ter5ao, where the measurements of $\dot{x}$ do constrain the orbital orientation, as described in section~\ref{subsec:Ter5ao}. However, for Ter5ao, the constraints on $\dot{x}_{\mu}$ were not taken into account in the Bayesian mass estimates, as they would require a 3-D map in $M_{\rm c}$, $\cos i$ and orbital orientation ($\Omega$). The results of such a map would be qualitatively similar, with the further exclusion of a small range of orbital inclinations close to $90 \, \deg$, but no additional constraints for lower inclinations. Quantitatively they would also be similar, since $\dot{x}$ is not measured with high significance. This means that for this pulsar there are no solid constraints on the individual masses. The deduced constraints on the individual masses and orbital inclination of Ter5ao, Ter5ap, Ter5au, and Ter5aw are summarised in Table \ref{tab:Ter5_mass_estimates}. 

\begin{table*}[h]
\caption[]{Constraints on the orbital inclination $i$, pulsar mass ($\Mp$), companion mass ($\Mc$), and the total mass ($\Mtot$) using the $\chi^2$ map technique for Ter5ao, Ter5ap, Ter5au, and Ter5aw for different confidence limits (C.L.). Ter5au has a nearly flat pdf in $i$ and hence no corresponding constraints.}
\centering
\renewcommand{\arraystretch}{1.8}
\begin{tabular}{|l|l|l|l|l|l|l|l|l|}
\hline
\hline
Pulsar & \multicolumn{2}{|c|}{J1748$-$2446ao} & \multicolumn{2}{|c|}{J1748$-$2446ap} & \multicolumn{2}{|c|}{J1748$-$2446au} & \multicolumn{2}{|c|}{J1748$-$2446aw}  \\ 
\hline
C.L. & 68.3\%  & 95.4\%  & 68.3\%  & 95.4\%  & 68.3\%  & 95.4\% & 68.3\%  & 95.4\% \\  
\hline
$i$ (deg.)      &  $67.7^{+9.9}_{-20.6}$ & $ 67.7^{+14.9}_{-41.7}$ & $71^{+12}_{-16}$ & $71^{+18}_{-30}$  &    -  & -  & $75.6^{+7.3}_{-14.}$ & $75.6^{+11.5}_{-37.8}$\\
$\Mp$ ($\msun$)    &  $2.16^{+0.06}_{-0.27}$ & $2.16^{+0.09}_{-1.16}$  & $1.700^{+0.015}_{-0.045}$ & $1.700^{+0.023}_{-0.134}$ & $1.45^{+0.09}_{-0.18}$ & $1.45^{+0.16}_{-0.75}$& $1.32^{+0.47}_{-0.31}$& $1.32^{+0.98}_{-0.81}$\\
$\Mc$  ($\msun$)  &   $0.99^{+0.27}_{-0.05}$ & $0.99^{+1.15}_{-0.07}$ & $0.294^{+0.046}_{-0.014}$ & $0.294^{+0.136}_{-0.017}$ & $0.35^{+0.19}_{-0.04}$& $0.35^{+0.77}_{-0.06}$ & $0.84^{+0.19}_{-0.16}$& $0.84^{+0.56}_{-0.31}$\\
$\Mtot$ ($\msun$) &   $3.166 \pm 0.024$   & $3.166 \pm 0.047$  & $1.997 \pm 0.006$ & $1.997 \pm 0.013$ & $1.82 \pm 0.07 $& $1.82 \pm 0.14$& $2.16^{+0.62}_{-0.56}$& $2.16^{+1.29}_{-1.06}$\\
\hline
\hline
\end{tabular}
\label{tab:Ter5_mass_estimates}
\end{table*}

\section{Discussion and future prospects}
\label{sec:discussion}

The ten new discoveries made in Ter5 not only show a wide range of properties but also raise open questions, particularly pertaining to formation and stellar evolution channels.

\subsection{Eccentric binaries}
\label{subsec:eccentric_binaries}

From the large measured $\omegadot$, we know that Ter5ao possesses a large binary mass ($3.166 \pm 0.024 \, \msun$), a large minimum companion mass ($M_{\rm c,min} = 0.93\, \msun$) and significant eccentricity ($e=0.32$). In the Galactic field, the most likely possibility would be that the system is a double neutron star binary, where a second supernova explosion from the progenitor of the companion has induced the observed eccentricity \citep{Tauris_2017}. However, given the pulsar's spin period ($P = 2.27$ ms), small intrinsic $\Pdot$ ($1.13 \times 10^{-20} \rm s\, s^{-1}$) and small  magnetic field ($1.6 \times 10^8\,$G), a prolonged episode of mass accretion from a low-mass companion via Case B RLO \citep{Tauris_2011} would be required, or potentially Case A \citep{Tauris_Langer_Kramer_2011}. This implies that the companion star was relatively light and gradually evolved into a low-mass WD in a circular orbit, as seen for most such systems in the Galaxy \citep[see e.g. review by][]{Tauris_2011, Tauris_Langer_Kramer_2011, Tauris_van_den_Heuvel_2023}. 

However, because of the large stellar encounter rate in Ter5 - in particular the large stellar encounter rate per binary ($\gamma$) that \cite{Verbunt_Freire_2014} predicted for this GC - 
it is possible that this low-mass WD companion was replaced by a more massive degenerate object in a secondary exchange encounter. Given the chaotic nature of this process, it almost invariably results in highly eccentric orbits \citep[see Section 5 in][]{Phinney_1992}. Many such eccentric and high mass companion systems already exist in Ter5 like J, Q, U and ai (Ransom et al., in prep) as well as  globular clusters with a high $\gamma$ \citep[see e.g.][and references therein]{Balakrishnan_2023}.

Given this possible origin of the companion of Ter5ao, we cannot infer anything about its nature based on considerations of stellar evolution. This degenerate companion could be another NS; if confirmed, it would make Ter5ao simultaneously the fastest spinning pulsar in any known DNS system and the widest orbit for any known DNS. This could be confirmed either with the measurement of a large mass, or the detection of radio pulsations from the companion of Ter5ao, the latter of which is currently being investigated. The hypothesis of an equal-mass DNS implies a relatively low orbital inclination of $35\, \deg$.

If the system is closer to edge-on ($i \sim 90 \, \deg$), then the mass of the pulsar could be up to $2.23 \, \msun$, and in this case the companion would have a mass of only $0.93 \, \msun$ - very likely a massive WD. This pulsar mass would be larger than the largest well-measured mass of a neutron star previously obtained from PSR~J0740+6620 \citep[$2.08 \pm 0.07\, \msun$;][]{Fonseca_2021} and would serve as an excellent test bed for constraining the equation of state of super-dense matter \citep{Ozel_Freire_2016}. However, such high orbital inclinations would result in a detectable Shapiro delay. Our current measurement is not precise enough to claim such a detection. Currently we do not have a significant detection of $\xdot$ for Ter5ao. We note, however, that a detection of a large value of $\xdot$ from future observations could exclude large orbital inclinations.

For Ter5ap, we can introduce tighter constraints on the individual masses, from the non-detection of the Einstein delay $\gamma_{\rm E}$. The pulsar mass $1.70^{+0.02}_{-0.13}\, \msun$ (95.4\% C. L.) is the largest measured for a pulsar in a globular cluster, the previous one being PSR J1910$-$5959A ($1.556^{+0.067}_{-0.076} \, \msun$; \citealt{Corongiu_2023_NGC6752A}).
An exchange encounter is also a possibility for the formation of this highly eccentric ($e = 0.905$) system. Among recycled pulsars, the eccentricity of this system is only second to NGC~6652A \citep{DeCesar_2015}. The latter authors suggested that the most suitable explanation for the high eccentricity and massive companion ($M_{\rm c, min} = 0.73\, \msun$) of NGC~6652A is the origin of the system in an exchange encounter, as discussed in the case of Ter5ao. However, for Ter5ap, the companion is much lighter ($M_{\rm c} = 0.294\, \msun$).
 This means that the latter system could either have formed in an exchange encounter, or it could have maintained its original He WD companion, with the eccentricity being later raised by gravitational perturbations from encounters with nearby stars. The time required to induce a certain eccentricity (assuming $e~>~0.01$) can be quantified as \citep[see Equation 5 in][]{Rasio_Heggie_1995}  
\begin{equation}
\label{eqn:time_eccentricity_high}
t_{>e} = 2 \times 10^{11} {\rm yr} \, \left(\frac{n}{10^4\, \rm pc^{-3}}\right)^{-1} \left(\frac{v}{10\, \rm km\, s^{-1}}\right) \left(\frac{\Pb}{\rm d}\right)^{-2/3}\, [\mathrm{-ln}(e/4)]^{-2/3},
\end{equation}
where $n$ is the number density of stars near the pulsar, $v$ is the one dimensional velocity dispersion in the core, $P_b$ is the orbital period and $e$ is the observed eccentricity. We take $v = 15.6\, \rm km \, s^{-1}$ for Ter5 from \cite{Baumgardt_Vasiliev_2021}. In order to derive $n$ we make assumptions similar to \cite{Lian_2023}. They used $n \propto \rho_{c}$ where $\rho_{c}$ is the core luminosity density of the globular cluster. Using the values for NGC 5024 as a reference (as given in \cite{Lian_2023} for convenience) and $\rho_c = 1.38 \times 10^{5}\,\rm L_{\sun}\, pc^{-3}$ for Ter5 as given in \cite{Harris_2010}, we get $n = 3.3 \times 10^{5}\,\rm pc^{-3}$. Applying these values to Ter5ap, we can estimate the time it would take for close encounters to raise the eccentricity to this value: $t_{>e, \rm ap} \sim 0.94 \, \rm Gyr$, which fits well within the age of the cluster \cite[12 Gyr; ][]{Ferraro_2016}. 

\subsection{Less eccentric binaries}

Apart from Ter5ao and ap, the other Ter5 discoveries show a range of eccentricities from essentially zero for the eclipsing systems to 0.025 for Ter5au. These small eccentricities indicate that none of these companions were exchanged, and because of this we can make some inferences about the nature of these systems from basic considerations of stellar evolution.

For instance, from the mass functions, and total mass measurements, we can say that Ter5au and Ter5aw likely have CO WD companions. As discussed above, the orbital and spin periods make the Ter5au system remarkably similar to PSR~J1614$-$2230, which likely evolved through case A RLO \citep{Tauris_Langer_Kramer_2011}. Other systems like PSRs~J1125$-$6014 \citep{Shamohammadi_2023} and J1933$-$6211  \citep{Geyer_2023} are very similar and likely had a similar origin. The only difference about Ter5au is the much larger eccentricity, likely caused by its location in a dense GC. A measurement of the individual masses of Ter5au would be important to confirm the nature of the companion mass. For Ter5aw, the companion mass is consistent with other CO WD companion pulsar binaries known in the Galactic field \citep{Mckee_2020, Shamohammadi_2023}. Moreover, this system is remarkably similar, in its companion mass and orbital and spin periods, to PSR~J1952+2630 \citep{Gautam_2022b}.

Given their lower masses, Ter5av and ax likely have He WD companions, with Ter5av likely seen at a low orbital inclination. None of these systems shows eclipses.
Their eccentricities (in particular, that of Ter5ax, $e \sim 0.009$) are still significantly larger (in the case of Ter5av, by three orders of magnitude) than those predicted by the relation between eccentricity and orbital period for He WD companions (Equation 7.35 from \citealt{Phinney_1992} also known as the Phinney relation).
This eccentricity could be induced from exchange encounters or from close flybys of nearby stars as discussed in Section \ref{subsec:eccentric_binaries}. For the latter case and assuming low eccentricities ($e~<~0.01$), Equation \ref{eqn:time_eccentricity_high} is slightly modified  as \citep[see e.g.][]{Rasio_Heggie_1995, Lian_2023}
\begin{equation}
\label{eqn:time_eccentricity_low}
t_{>e} = 4 \times 10^{11} {\rm yr} \, \left(\frac{n}{10^4\, \rm pc^{-3}}\right)^{-1} \left(\frac{v}{10\, \rm km\, s^{-1}}\right) \left(\frac{\Pb}{\rm d}\right)^{-2/3}\, e^{2/5}.
\end{equation}

For Ter5ax, we use Equation \ref{eqn:time_eccentricity_low} to derive $t_{>e, ax} \sim 0.3\, \rm Gyr$. This is consistent with  the estimated age of Ter5 \citep[12 Gyr, ][]{Ferraro_2016}. For the other He WD system, Ter5av, we get a similar value $t_{>e, av} \sim 0.3\, \rm Gyr$.         
A similar argument can also be applied to Ter5au (which also has a large $e$ compared to pulsar - CO WDs observed in the disk), this yields $t_{>e, au} \sim 1.31 \, \rm Gyr$, also consistent with the age of the cluster. Ter5av has a relatively lower eccentricity but still higher than expected from the Phinney relation \citep{Phinney_1992}.    

\begin{figure}
    \centering
    \includegraphics[width=0.48\textwidth]{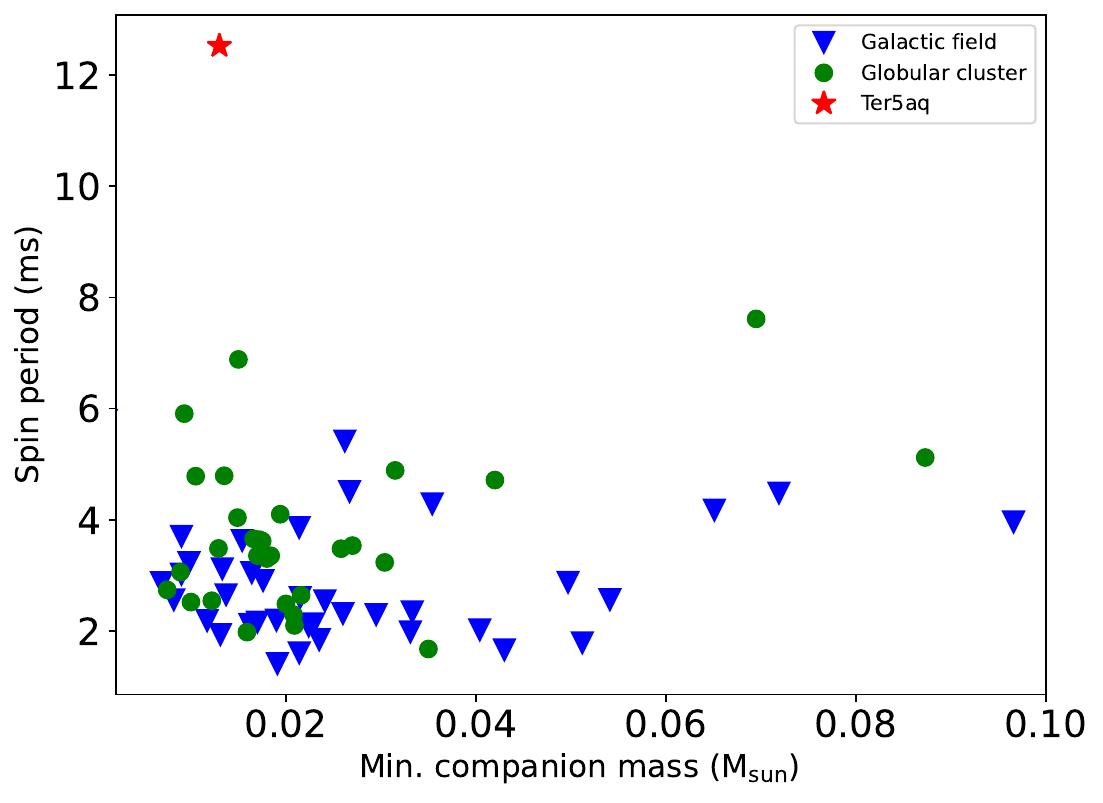}
    \caption{Spin period plotted as a function of the minimum companion mass for all known black widow systems. No clear distinction is seen between these pulsars whether they are in the Galactic field or within globular clusters. Ter5aq stands out given its relatively slow spin period amongst the rest of the population.}
    \label{fig:Ter5aq_p_mcmin}
\end{figure}

Among the discoveries showing similar properties to spider systems, Ter5aq ($P = 12.52\, \rm ms$) seems to be an outlier. Figure \ref{fig:Ter5aq_p_mcmin} shows the spin period vs minimum companion mass for all known black-widow pulsars. The spin period of Ter5aq is much slower than the rest of the black widow population including the Galactic as well as globular cluster pulsars. Although there are relatively slow spinning redback systems like Ter5A ($P = 11.56\, \rm ms$), the relatively lower companion mass of black-widow systems suggests that the mass loss is higher and the pulsar could be more recycled and hence spin faster in black-widow systems. However, given that the companion material could undergo evaporation, this may not always be true. Given that there is no constraint on the intrinsic spin period derivative, it is currently difficult to comment further on the recycled nature of Ter5aq.

\begin{figure}
    \centering
    \includegraphics[width=0.48\textwidth]{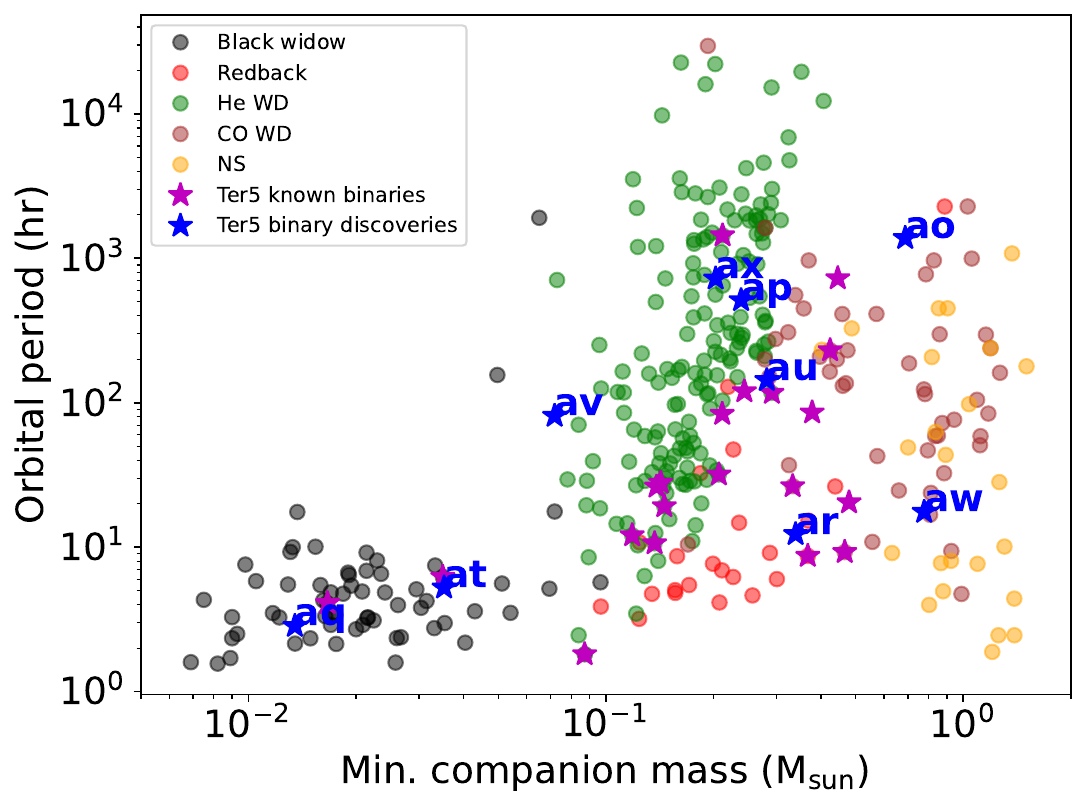}
    \caption{Orbital period plotted as a function of minimum companion mass for all binary pulsars. The information was obtained from the ATNF catalogue \citep{Manchester_2005}. The different colours represent the different types of companions. All Ter5 binaries discovered in this work are marked in blue and their respective letters are also indicated.}
    \label{fig:Ter5_all_pb_mc}
\end{figure}

While Ter5ar and Ter5at are most likely spider systems especially given their eclipsing nature, the nature of the companion of Ter5av is debatable. Firstly, we do not observe eclipses with Ter5av. Despite this, if assumed to be a spider with a high orbital inclination, the minimum companion mass ($M_{\rm c,min} \sim 0.07\, \msun$) places it in between the black widow and redback populations. However, the relatively high orbital period of Ter5av suggests it is not a spider system but instead has a He WD companion (see Figure \ref{fig:Ter5_all_pb_mc}) with a low inclination angle.

Ter5at and Ter5ar are eclipsing black widow and redback systems respectively whose spin period, orbital period and derived minimum companion masses are consistent with the known spider population (see Figure \ref{fig:Ter5_all_pb_mc}). The discovery of Ter5ar and its corresponding association with VLA-38 from \cite{Urquhart_2020} as well as the matching orbital period between radio and X-ray data  demonstrates the importance of continued synergies between imaging and time-domain searching through a multiwavelength approach. Furthermore, identifying steep spectrum sources in radio imaging similar to \cite{Urquhart_2020} are particularly useful with interferometers like MeerKAT where narrow synthesised beams could be placed on these positions. Besides this, apart from Ter5ar, three other discoveries show likely associations with X-ray sources. All except Ter5av show eclipsing properties further cementing the well known association of spider systems with bright X-ray sources, as mentioned earlier. It also  provides a platform to better understand the properties of these systems from a multi-wavelength approach. 

\subsection{The pulsar population of Ter5}

As a collective, the ten new discoveries offer scope on understanding several aspects of globular cluster properties. Nine out of the ten are binary systems making the overall fraction of binaries in Ter5 as 59\%.  This is the sixth highest percentage among all globular clusters with more than 10 known pulsars following M62, NGC~362, M28, 47~Tuc, and NGC~1851. M62 (which has 10 known pulsars) is a unique case where all known pulsars are in binaries. Moreover, the evolution history and core-collapsed nature of M62 is still a matter of open debate \citep{Vleeschower_2024}. NGC~362 and NGC~1851 are known to have compact cores thus making the high fraction of binaries less surprising. On the other hand, Ter5 is established as a bulge cluster similar to M28 \citep[e.g.][]{Ferraro_2009}. \cite{Ferraro_2016} suggested that Ter5 likely underwent multiple episodes of star formation in its history given the trimodal distribution of stars based on iron content. This in turn could explain why no other globular cluster currently possesses more confirmed pulsars than Ter5. Moreover, the high mass and density of Ter5 contribute to a high stellar encounter rate which in turn encourages a large fraction of binary pulsars to be formed. It is also worth noting that selection effects from the types of searches conducted so far play a major role in the binary fraction. For example, at least five isolated MSPs in Ter5 have been discovered by specifically applying stack searches (Ransom et al., in prep). Such a method applied to other globular clusters could also boost their respective isolated pulsar numbers.  

\cite{Prager_2017} already undertook a study using 34 pulsars to help place constraints on a possible black-hole in the core of Ter5 and better constrain the structural properties of the cluster. Their results favoured the argument that Ter5 is a fragment of the Galactic bulge rather than a remnant of a dwarf galaxy. Given the addition of 15 more pulsars since, it is worth conducting a similar study to improve the constraints and confirm these findings. Finally, a large collection of pulsars in Ter5 could also potentially probe the presence of ionised gas in the intra-cluster medium, similar to the work done on 47 Tuc by \cite{Freire_2001b}. However, a similar study for Ter5 in the past has yielded ambiguous results owing to high DM in the foreground unlike 47 Tuc. Nevertheless, multiple pulsars at multiple DMs and positions can better ascertain the presence or absence of ionised gas within the cluster.  \cite{Martsen_2022} already obtained precise rotation measure (RM) values from 28 pulsars to deduce constraints on the parallel component of the magnetic field component along the line of sight to Ter5. Polarisation studies of the latest discoveries could help further update these constraints.

\subsection{Prospects}
\begin{figure}
    \centering
    \includegraphics[width=0.48\textwidth]{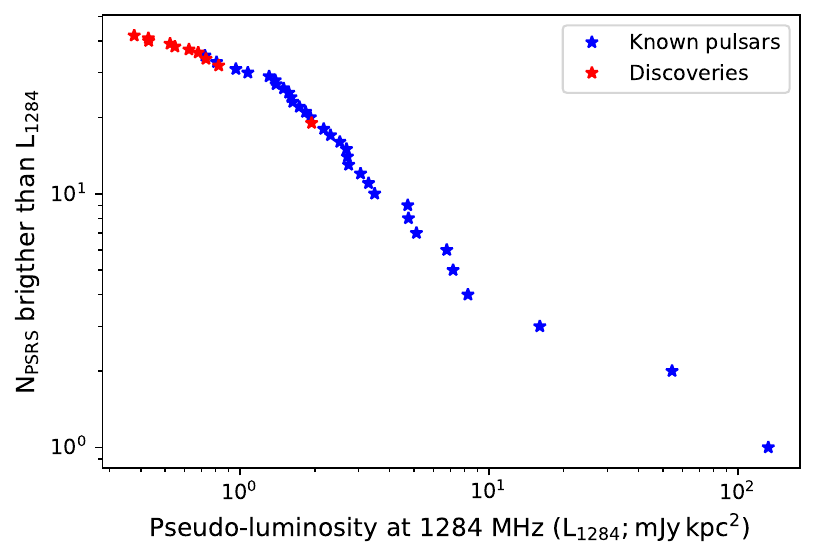}
    \caption{Cumulative number of pulsars with pseudo-luminosity greater than a given pseudo-luminosity ($\rm L_{1284}$) as a function of the pseudo-luminosity. This figure is adapted from Figure 3 in \citet{Martsen_2022}. We have used the known pulsar values as quoted in \citet{Martsen_2022} and added the radiometer flux based estimates (see Section \ref{subsec:flux_density}) for the ten new pulsars. It is evident that the linear trend tapers when going from right to left.}
    \label{fig:logN_logS}
\end{figure}

There are various reasons to continue searching for pulsars in Ter5. Firstly, multiple simulations using different methods have predicted a larger population of pulsars yet to be discovered. \citet{Bagchi_2011} conducted Monte-Carlo simulations and modelled the observed population of pulsars as the tail end of a broader intrinsic luminosity distribution function. They used a log-normal distribution and estimated $\sim$ 150 pulsars to be detectable in Ter5. \citet{Chennamangalam_2013} built on this work and used Bayesian statistics to constrain the luminosity function parameters. In the best case scenario, they predicted more than 200 detectable pulsars. Adding to this, Figure \ref{fig:logN_logS} is adapted from Figure 3 in \citep{Martsen_2022} and shows the number of pulsars brighter than a certain luminosity value as a function of the pseudo-luminosity. The figure has been updated with the luminosity values of the ten newly discovered pulsars and it still shows a turnover suggesting that we are approaching a minimum luminosity cutoff and there are potentially some more pulsars to be found. Its noteworthy that half of these discoveries are fainter than any previously known pulsars.   

Secondly, the searches conducted so far have multiple limitations. All the discoveries made with MeerKAT so far have resulted from searches limited to the beams within the core of the cluster alone. Searches in the outer beams especially within the half-mass and half-light radius still have the potential to boost discovery numbers. These pulsars in the outer regions of the cluster are expected to have lower mass companions if in binaries (owing to mass segregation) or isolated pulsars that have been flung out from interactions. It is also worth noting that previous searches have used several datasets to stack Fourier power spectra to boost the S/N of possible pulsar candidates \citep{Cadelano_2018}. While being effective for isolated pulsars, binary pulsars however would undergo significant drift in spin frequency leading to leakage of Fourier power over several bins and in turn reducing the effective S/N. It was thus assumed that any isolated pulsars hidden in Ter5 data would be recovered from the stack searches. However, the discovery of Ter5as proved otherwise. This is due to the significant drift in spin frequency with time (in turn reducing the stacked Fourier power) caused by the spin frequency derivatives. A significant fraction of discoveries with eclipsing properties also showed that there are possibly other pulsars yet to be seen owing to observing at unsuitable orbital phases.

The searches conducted on MeerKAT data so far have also been limited to acceleration searches down to 30 min segments. This implies that the searches are reasonably sensitive to binary pulsars with orbital periods of the order of hours rather than minutes. Applying jerk searches \citep{Andersen_Ransom_2018} or coherent template bank based searches \citep{Allen_2013, Balakrishnan_2022} can thus provide a possibility to find these highly compact binaries whose orbital period is of the order of few minutes. Such binaries can provide an excellent platform for testing GR in stronger gravitational fields than those of known binary pulsars. We are currently working with MeerKAT data to search for such systems  using Einstein@Home \citep{Anderson_2006}, a volunteer distributed computing project that has already been successful in the past for discovering several radio pulsars \citep[see][and references therein]{Knispel_2015}. 

Ter5 will also be observed in the near future with the recently installed S-Band (1.7-3.5 GHz) receiver system \citep{Kramer_2016_Sband,Padmanabh_2023} at MeerKAT. This frequency band has historically been successful with the GBT primarily owing to the lower impact of dispersion compared to L-Band. The enhanced sensitivity of MeerKAT would only boost the possibility of finding several more pulsars in Ter5. 

Thirdly, the detection of systems like Ter5ao, which are clearly the products of secondary exchange interactions, implies that other such systems might be found in Ter5. These could include MSP--MSP binaries, or even MSP--stellar mass black hole binaries, which would represent new laboratories for tests of gravity theories.

Finally, the long timing baselines of the ten new pulsars has only been possible from the use of archival data from the GBT. Thus any further discoveries made would also benefit from the already existing rich archival dataset for obtaining quick and robust timing solutions in the future.             

\section{Conclusions}
\label{sec:conclusions}

We have discovered ten new pulsars using the MeerKAT telescope in the Ter5 globular cluster as part of the TRAPUM globular cluster survey. This has brought the total number of known pulsars in this cluster to 49, the highest for any globular cluster known. We also presented long-term timing solutions nearing two decades for nine of these pulsars mainly using archival GBT data. These include astrometric parameters like proper motion as well as PK parameter measurements including $\Pbdot$ that helped derive intrinsic spin period derivative values for several pulsars.

Ter5ao is an eccentric, wide-orbit pulsar with a large minimum companion mass and a rapid spin period, suggesting that it is the result of a secondary exchange encounter. We were able to detect $\omegadot$ for this system, which yields a total mass of $3.17 \pm 0.04\, \msun$ to 95\% C. L. The system is either a DNS, or has a massive WD companion. The pulsar could be quite massive ($> 2\, \msun$), but at the moment we have no constraints on the individual component masses. Ter5ap has the second highest eccentricity for any recycled pulsar known ($e = 0.905$) and the highest mass measured for a pulsar in a globular cluster, but unlike other similar systems, its companion is relatively light. For this reason, we cannot conclude that it is the result of an exchange encounter. Two systems, Ter5aq and Ter5at, are black widow systems confirmed by the detection of radio eclipses. Ter5ar is an eclipsing redback system which is associated with the radio counterpart to the source named as VLA-38 by \cite{Urquhart_2020}. Ter5as is the lone isolated pulsar discovery among the ten pulsars. Ter5au, Ter5ap, Ter5av, and Ter5ax likely have WD companions, but their orbital eccentricities have most likely been acquired from gravitational perturbations from the dense surrounding environment within the core of Ter5.  We have a potential Shapiro delay in Ter5aw (currently with 2.4-$\sigma$ significance) leading to individual mass measurements of $\Mp = 1.32^{+0.47}_{-0.31} \, \msun$ and $\Mc = 0.84^{+0.19}_{-0.16} \, \msun$ to 68.3\% C. L., suggesting a rare CO WD companion. The wide variety of pulsars found in Ter5 demonstrates the richness of this cluster, not only to our understanding of stellar evolution but also for probing the intricacies of globular cluster dynamics and environments.

\begin{acknowledgements}
The MeerKAT telescope is operated by the South African Radio Astronomy Observatory, which is a facility of the National Research Foundation, an agency of the Department of Science and Innovation. SARAO acknowledges the ongoing advice and calibration of GPS systems by the National Metrology Institute of South Africa (NMISA) and the time space reference systems department of the Paris Observatory. MeerTime data is housed on the OzSTAR supercomputer at Swinburne University of Technology. PTUSE was developed with support from  the Australian SKA Office and Swinburne University of Technology. The National Radio Astronomy Observatory is a facility of the National Science Foundation operated under cooperative agreement by Associated Universities, Inc. 
The Green Bank Observatory is a facility of the National Science Foundation operated under cooperative agreement by Associated Universities, Inc. The authors also acknowledge MPIfR funding to contribute to MeerTime infrastructure. TRAPUM observations used the FBFUSE and APSUSE computing clusters for data acquisition, storage and analysis. These clusters were funded and installed by the Max-Planck-Institut für Radioastronomie and the Max-Planck-Gesellschaft.
PVP, PCCF, AR, FA, EDB, WC, DJC, CJC, AD, MK, VVK acknowledge continuing valuable support from the Max-Planck Society.
SMR is a CIFAR Fellow and is supported by the NSF Physics Frontiers Center award 1430284. MED acknowledges funding from the National Science Foundation Physics Frontier Center award No. 2020265. Pulsar research at UBC is supported by an NSERC Discovery Grant and by the Canadian Institute for Advanced Research. LZ acknowledges financial support from ACAMAR Postdoctoral Fellowship and the National Natural Science Foundation of China (Grant No. 12103069). The research activities described in this paper were carried out with contribution of the NextGenerationEU funds within the National Recovery and Resilience Plan (PNRR), Mission 4 - Education and Research, Component 2 - From Research to Business (M4C2), Investment Line 3.1 - Strengthening and creation of Research Infrastructures, Project IR0000034 – “STILES - Strengthening the Italian Leadership in ELT and SKA”. VVK acknowledges financial support from the European Research Council (ERC) starting grant ,"COMPACT" (Grant agreement number 101078094), under the European Union's Horizon Europe research and innovation programme. MBa acknowledges support through ARC grant CE170100004. BWS acknowledges funding from the European Research Council (ERC) under the European Union’s Horizon 2020 research and innovation programme (grant agreement No. 694745). AP, AR and MBu gratefully acknowledge financial support by the research grant `iPeska' (P.I. Andrea Possenti) funded under the INAF national call Prin-SKA/CTA approved with the Presidential Decree 70/2016. AP also acknowledges the support from the “Italian Ministry of Foreign Affairs and International Cooperation”, grant number ZA23GR03. LV acknowledges financial support from the Dean’s Doctoral Scholar Award from the University of Manchester. Finally, we thank the anonymous referee for their suggestions to bring clarity to the text.
\end{acknowledgements}

%
\bibliographystyle{aa} 
\bibliography{Ter5_discoveries} 
%
\end{document}